\documentclass[aps,twocolumn,amsmath,showkeys,showpacs,prl]{revtex4-1}
\usepackage{dcolumn}% Align table columns on decimal point
\usepackage{bm}% bold math
\usepackage[T1]{fontenc}
\usepackage{amsmath}
\usepackage{graphicx}
\usepackage{tabularx}

\usepackage{hyperref}
\hypersetup{colorlinks=true, linkcolor=blue, citecolor=blue, urlcolor=blue, pdftitle={Structural and electronic phenomena at oxyfluoride KTaO$_3$/K$M$F$_3$ ($M$ = Zn and Ni) superlattices: Rashba splitting and 2DEG}, pdfauthor={A. C. Garcia-Castro}}

\begin{document}
\title{Structural and electronic phenomena at oxyfluoride KTaO$_3$/K$M$F$_3$ ($M$ = Zn and Ni) superlattices: Rashba splitting and 2DEG} 
\author{A. C. Garcia-Castro$^{1}$}
\email{acgarcia@uis.edu.co}
\author{R. Ospina$^{1}$}
\affiliation{$^1$School of Physics, \textsc{cimbios} Research Group, Universidad Industrial de Santander, Bucaramanga, 680002, Colombia.}
%\email{acgarcia@uis.edu.co}
\author{Philippe Ghosez$^{2}$}
%\email{philippe.ghosez@ulg.ac.be}
\author{Eric Bousquet$^{2}$}
\email{eric.bousquet@uliege.be}
\affiliation{$^2$Physique Th\'eorique des Mat\'eriaux, Universit\'e de Li\`ege, B-4000 Sart-Tilman, Belgium.}
\author{Aldo H. Romero$^{3,4}$}
\email{alromero@mail.wvu.edu}
\affiliation{$^3$Department of Physics and Astronomy, West Virginia University, WV-26506-6315, Morgantown, USA}
\affiliation{$^4$Facultad de Ingenier\'ia-BUAP, Apartado Postal J-39, Puebla, Pue. 72570, M\'exico.}

\begin{abstract}
Here, we present the theoretical analysis of the structural and electronic degrees of freedom of two different oxide/fluoride perovskite superlattices, KTaO$_3$/KZnF$_3$ and KTaO$_3$/KNiF$_3$. Using first-principles calculations, we found the appearance of a two-dimensional electron, 2DEG, and hole, 2DHG, gases as a function of the number of layers of the different pristine materials. We demonstrate that the phonon-dynamics at the KTO/K$M$F superlattices play a crucial role in the appearance of these effects. 
Additionally, our results reveal a rather sizeable Rashba-type spin-splitting at these interfaces in comparison to another oxide/oxide counterparts.
\end{abstract}

\pacs{75.85.+t, 31.15.A-, 71.15.Mb, 75.50.-y, 77.65.-j}
\maketitle

\section{Introduction}
Unique electronic properties at oxide/oxide interfaces are accomplished by tuning the ferroic and spin properties of the chosen constituent materials. 
This optimization is driven by the complex electron-electron interactions in strongly correlated material's interfaces, which gives rise to a broad set of intriguing phenomena \cite{Hwang2012,Granozio2013}. 
In oxide/oxide interfaces, these phenomena include the appearance of two-dimensional electron gases (2DEG), and its counterpart, two-dimensional hole gases, 2DHG at the oxide insulator/insulator interfaces \cite{MRS:7956955,Joshua2012}, exotic electronic states due to strong spin-orbit coupling with an intriguing spin-texture \cite{PhysRevLett.109.196401}, metal-insulator transitions tunable by electric fields \cite{Zubko2011}, interfacial magnetic phases \cite{Nanda2010}, improper ferroelectricity \cite{Bousquet2008, Benedek2011, ADFM:ADFM201300210}, and multiferroic interactions \cite{Zanolli2013}. 
While studies of these phenomena have focused only on complex oxide materials, there are strong indications that complex fluorides may have analogous properties with respect to the oxide materials. 
In particular, our recent computational work on fluoride perovskites ($AB$F$_3$) indicates that interfaces between $AB$F$_3$ and other fluorides, or hybrid systems with isostructural oxide perovskites $AB$O$_3$, have unique advantages in terms of their magnetic, magneto-transport, and multiferroic properties \cite{PhysRevB.89.104107, PhysRevLett.116.117202, Yang2017}. 
To date, little experimental and theoretical work has been devoted to fluorides, either in bulk \cite{PhysRevB.85.224430,Dobson2011171, PhysRevB.39.9738, PhysRevB.69.033102} or in thin-films forms \cite{Pang2004, Yang2017} and therefore, there is minimal knowledge about their surfaces and interfacial properties \cite{Yang2017}. 
In addition to the studies of bulk fluorides, much less is known about the multifunctional properties of interfaces and surfaces of complex fluorides \cite{Yang2017}. By searching in the literature, it can be found that there are just a few reported examples of complex fluorides thin films and the corresponding interface characterization. These include Fe/KFeF$_3$ bilayers \cite{Pang2004} and the trilayered system Fe/KNiF$_3$/FeF$_2$ \cite{Widuch2011} that has a dynamic and rotatable exchange anisotropy. Such systems exhibit exchange bias, which results in a center shift of the hysteresis loop of the ferromagnetic Fe layer away from H = $0$ T. The latter as a result of the magnetic interface interaction with the antiferromagnetic material. Additionally, it has been shown that NaMnF$_3$ epitaxial thin-films grown onto SrTiO$_3$ exhibit ferroelectric and magnetic behavior thanks to the fluoroperovskite film \cite{Yang2017}.

Recently, several efforts have been dedicated to study and understand mixed systems by considering the effect of oxygen replacement by fluorine ions in oxide perovskites with different stoichiometries. Thus, a process called ``fluorination'' has been successfully applied to the epitaxial growth of SrFeO$_{3-x}$F$_x$ ($x$ from $0$ to $1$) thin films \cite{doi:10.1021/ja410954z} where substantial changes in the electric conductivity and transport properties have been demonstrated. This mainly due to an induced change in the nominal oxidation state of the Fe cation. Then, and even more surprising, multifunctional properties such as superconductivity in Sr$_2$CuO$_2$F$_{2+\delta}$ \cite{AI-Mamouri1994}, ionic conductivity at BaFeF$_x$O$_{3-\delta}$ \cite{doi:10.1021/cm102724k}, and robust antiferromagnetism in Ba$_{0.8}$Sr$_{0.2}$FeF$_x$O$_{3-\delta}$\cite{doi:10.1021/ja2028603} can be tuned and/or modified in fluorinated oxides ---also called oxyfluorides---.

On the other hand, mixed SrTiO$_3$ and KAgF$_3$, AgF$_2$ interfaces were theoretically proposed where, superconducting behavior has been predicted \cite{Yang2014,Yang2015}. Thus, oxyfluoride materials have been highlighted as materials with potential applications in the next-generation rechargeable batteries, with enhanced properties compared to their oxide counterparts \cite{CNMA:CNMA201600342}. Moreover, a geometrically-driven ferroelectric instability \cite{PhysRevB.89.104107} was stabilized under strain in the NaMnF$_3$ fluoride thin-films epitaxially grown on SrTiO$_3$ \cite{PhysRevLett.116.117202, Yang2017} as commented before. Besides, heteroanionic materials (\emph{i.e.}, oxyfluorides, oxynitrides, and oxysulfides) are recently on the focus of debate due to their potential applications in novel materials by design \cite{Charles2018, Kageyama2018, Harada2019}.
In general, these results make of these systems an appealing field with high potential, but still weakly explored.

In this paper, we report a first-principles investigation of fluoride/oxide perovskite interfaces and superlattices of KTaO$_3$/K$M$F$_3$ (M = Zn and Ni) compositions. After full geometric and electronic optimization, we demonstrate the appearance of a 2DEG and 2DHG. Also, the analysis of their properties in terms of an electrostatic model, across the oxyfluoride superlattices, are presented. The latter, followed by the analysis of the magnetic structure for $M$ = Ni. 
Finally, a Rashba-type spin-splitting at the interface is examined and discussed.
With this study, we hope to lead to novel strategies for searching new and complex perovskite-based oxyfluoride heterostructures.

%%%%%%%%%%%%%%%%%%%%%%%%%
\section{Computational Details}
We performed first-principles calculations within the density-functional theory, DFT \cite{PhysRev.136.B864, PhysRev.140.A1133}, as implemented in the Vienna $Ab$-initio Simulation Package (\textsc{vasp} 5.3.3) \cite{Kresse1996, Kresse1999}. 
We used projected augmented wave (PAW) method \cite{Blochl1994} to represent the valence and core electrons. The electronic configurations taken into account as valence electrons were: K (3$p^6$4$s^1$, version 17Jan2003), Ta (5$p^6$5$d^5$6$s^2$, version 07Sep2000), Zn (3$d^{10}$4$s^2$, version 06Sep2000), Ni (3$p^6$3$d^8$4$s^2$, version 06Sep2000), F (2$s^2$2$p^5$, version 08Apr2002), and O (2$s^2$2$p^4$, version 08Apr2002). The exchange-correlation was represented within the generalized gradient approximation (GGA) with the PBEsol flavor \cite{Perdew2008}. 
This has been proven to be one of the most accurate XC functional for treating oxyfluoride compounds \cite{PhysRevB.94.174108}. 
The magnetic character was considered, and the $d$-electrons were corrected by means of the DFT+\emph{U} within the Liechtenstein formalism \cite{Liechtenstein1995}. The \emph{U} value was found by fitting the electronic gap and the magnetic moment with respect to those obtained by HSE06 hybrid-functionals \cite{HSE, HSE06} in the 1/1 superlattice. The periodic solution of these crystalline structures was represented by using Bloch states with a  $\Gamma$-centered \emph{k}-point mesh of 6$\times$6$\times$\emph{x} (\emph{x} dependent on the $c$ parameter of the supercell) and 700 eV energy cut-off, which has been tested already to give forces convergence to less than 0.001 eV$\cdot$\r{A}$^{-1}$.
Spin-orbit coupling, SOC, was included such that its effect into the electronic structure could be analyzed \cite{Hobbs2000}. Born effective charges and phonon calculations were performed within the density functional perturbation theory (DFPT) \cite{gonze1997} as implemented in the \textsc{vasp} code and analyzed through the Phonopy interface \cite{phonopy}.

To perform calculations based-on hybrid functionals for large supercells (above 2/2 ratio), and to correlate the results of the electronic structure with the DFT+$U$ results obtained with the \textsc{vasp} code, we have used the linear combination of atomic orbitals (LCAO) based \textsc{crystal} code \cite{Roberto2005}. 
This code uses a Gaussian-type basis set to expand the Kohn-Sham orbitals, including polarization effects, with the advantage of a smaller basis set is required. 
Additionally, all the electrons were included in K, Ni, Zn, O, and F, while for the Ta case, it was used a pseudopotential method. For the exchange-correlations term, the hybrid functionals taken into account are the B1WC \cite{PhysRevB.77.165107}, B3LYP \cite{:/content/aip/journal/jcp/98/2/10.1063/1.464304,:/content/aip/journal/jcp/96/3/10.1063/1.462066}, and PBEsol$0$ (derived from the PBE$0$ functional and based-on the PBEsol approximation) \cite{PBE0} all of them with a 16\%, 20\%, and 25\% of exact exchange-correlation term respectively. The latter in order to investigate the correlation and functionals' effect into the calculations.

%%%%%%%%%%%%%%%%%%%%%%%%%
\section{Results and Discussion}
In what follows, we discuss the appearance of 2DEG and 2DHG at the KTaO$_3$/K$M$F$_3$ ($M$ = Zn and Ni) oxyfluorides interfaces, the magnetic structure when $M$ = Ni and finally, the Rashba spin-splitting at the bands' interface.

\begin{figure*}[!t]
 \centering
 \includegraphics[width=14.5cm,keepaspectratio=true]{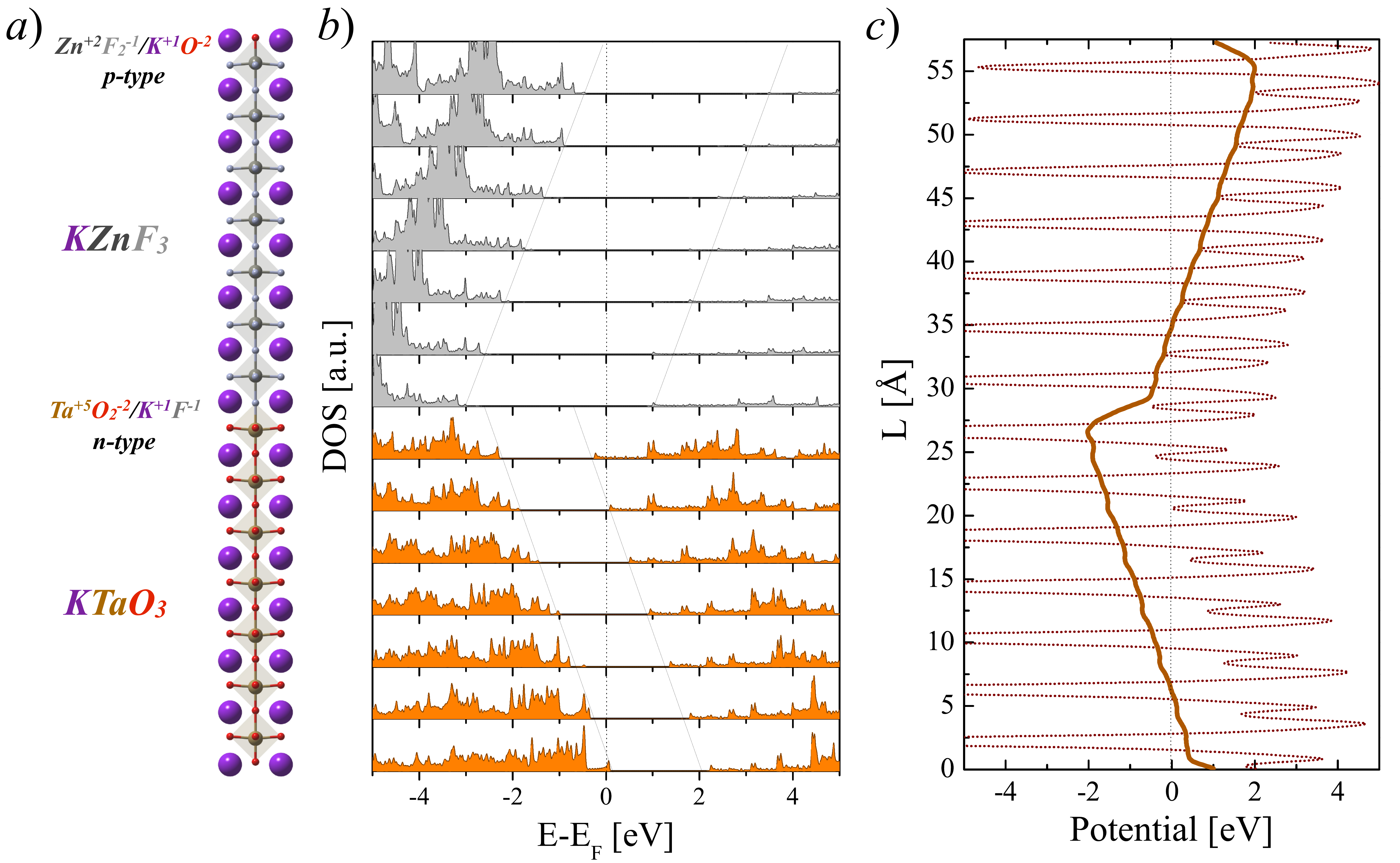}
 \caption{(Color online) $a$) 7/7 KTaO$_3$/KZnF$_3$ superlattice where it can be observed the $B$-sites of centering and the TaO$_2$/KF and ZnF$_2$/KO interfaces. $b$) Layered revolved DOS for in which a 2DEG and 2DHG are shown at the TaO$_2$/KF and ZnF$_2$/KO respectively. $c$) Computed electrostatic potential along the $z$-axis showing a zig-zag profile as expected for superlattices formed from alternating $n$ and $p$-type layers.}
 \label{fig:KTO-KZF-U0}
\end{figure*} 

%%%%%%%%%%%%%%%%%%%%%%%%%
\subsection{2DEG and 2DHG at the interfaces:}
At the bulk level, KTaO$_3$ ($a$= 3.989 \r{A}) and K$M$F$_3$ ($M$ = Zn and Ni with $a$ = 4.055 \r{A} and 4.012 \r{A} respectively) are insulator perovskites that belong to the cubic $Pm\bar{3}m$ symmetry \cite{ref1, Knox1961, doi:10.1143/JPSJ.14.1823, PhysRevB.50.8166}.
We studied the superlattices (KTaO$_3$)$_n$/(K$M$F$_3$)$_n$, where $n$ is the number of atomic layers (in unit cells, u.c.) in each compound. 
To do so, we considered $n$ values ranging from 1 u.c. to 8 u.c.  The $in$-plane lattice parameter was set to 4.022 \r{A} (4.001 \r{A})   for the KTO/KZF (KTO/KNF) superlattice. %\textcolor{red}{E: why these values specifically?}
These values correspond to the averaged bulk values between KTO/KZF and KTO/KNF.
Such $in$-plane values represent a strain of +0.84\% and $-$0.81\% for the KTO and KZF at the KTO/KZF superlattice and +0.30\% and $-$0.27\% for the KTO and KNF at the KTO/KNF.

The superlattice structure for the $n$ = 7 u.c. is schematically shown in Fig. \ref{fig:KTO-KZF-U0}$a$ with $M$ = Zn.
Here, due to the unbalanced charge at the Zn$^{+2}$F$_2^{-1}$/K$^{+1}$O$^{-2}$ and Ta$^{+5}$O$_2^{-2}$/K$^{+1}$F$^{-1}$ interfaces, an alternating \textit{n}- and \textit{p}-type interfaces can be appreciated. 
After full atomic and electronic relaxation, we observed that the K--K distance, (\emph{i.e.}, equivalent to the lattice parameter $c_i$ in which, $i$ is the layer index) is $c_1$ = 4.135 \r{A}, close to the ZnF$_2$/KO interface. Then, for the rest of the KTO layers, from 2 u.c. to 6 u.c., the $c_{2-6}$ is equal to 4.060 \r{A}.
Close to the TaO$_2$/KF interface we have $c_7$ = 4.231 \r{A}. 
As expected, the K--K distance for KTO at the interfaces is larger, mainly due to the O--Ta--F bonding. 
Interestingly, in the rest of the material, the $c$ parameter represents a $xy$-plane expansion strain, when compared to the $in$-plane lattice of $a$ = 4.022 \r{A}. 
We observe the same behavior for the KZF compound with $c_8$ = 4.118 \r{A}, $c_{9-15}$ =  4.110 \r{A}, and $c_{16}$ = 3.897 \r{A}. The latter lattice parameter is explained in terms of the additional contraction of the Zn-octahedra due to the F--Zn--O bonding. 
When looking and analyzing the electronic structure, as anticipated from the ``polar catastrophe'' scenario, an insulator-to-metal transition, IMT, as a function of the number of layers takes place \cite{PhysRevB.80.045425}, (see the layered-resolved DOS also in Fig. \ref{fig:KTO-KZF-U0}$b$). Consequently, we found the formation of the 2DEG and 2DHG in the KTaO$_3$/K$M$F$_3$ superlattices induced by the local electronic reconstruction for thicknesses $n$ > 6 u.c.  \cite{PhysRevB.80.045425}.  

This phenomenon occurred mainly due to the electrons and holes donated by each layer at the interface. The latter charge transfer is a consequence of the compensation of the electrostatic potential generated by the additional electrons in the system due to the charged layers in KTO. 
Added to the polar catastrophe phenomena, we could expect that the inhomogeneous charge distribution along the O--$B$--F bonding would have a contribution to the electronic reconstruction process. 
It is important to note that the ITM transition in oxyfluoride superlattices, at 6/6 ratio, is achieved with less polar/non-polar layers than in oxides systems such as LaAlO$_3$/SrTiO$_3$ in which, the IMT is observed for values above the 8/8 ratio \cite{PhysRevB.80.045425}.
Another point that is important to highlight (see Fig. \ref{fig:KTO-KZF-U0}$b$) is that both 2DEG and 2DHG, are mostly localized at the 5$d_{xy}$--Ta and 2$p$-O orbitals, respectively. 
Thus, the fluoride bands do not have a contribution close to the Fermi energy in the electronic structure, and the electrons and holes are localized only in the oxide film. 
Therefore, we believe that the fluoroperovskite film is acting as a ``condensing'' layer retaining the 2DEG and 2DHG at the oxide layers. 

Additionally, from the electrostatic point-of-view, a larger electric field across the entire system is found in both cases, in comparison to the LAO/STO oxide/oxide counterpart. For example, from the DOS analysis we found that when $M$ = Zn: $\mathcal{E}$ = 105 mV$\cdot$\r{A}$^{-1}$ (and $\mathcal{E}$ = 98 mV$\cdot$\r{A}$^{-1}$ when $M$ = Ni) against $\mathcal{E}$ = 58 mV$\cdot$\r{A}$^{-1}$ at LAO/STO \cite{PhysRevB.80.045425} this before the Zener breakdown. In the oxyfluoride superlattices, the same value of the electric field was obtained from the DFT computed electrostatic potential (in Fig. \ref{fig:KTO-KZF-U0}$c$) calculated along the $z$-axis. 
Here, a zig-zag profile is observed as expected for alternating $n$- and $p$-type interfaces. 
It worth mentioning that the thickness superlattice ratio experimentally expected for the Zener breakdown should be larger, close to 9/9. This due to the well-known DFT underestimation of the bandgap energy, which, in the KTaO$_3$, predicts 2.2 eV against 3.64 eV measured value \cite{PhysRevB.74.155130}.

At the metallic interface, we observe that the effective masses in both $M$ = Zn and Ni cases are close to 0.3$m_e$ for the electrons at the 2DEG that lies in the lowest 5$d_{xy}$:Ta band. In contrast, heavier hole carriers with $m^*$ of 1.1$m_e$ are found to be localized at the 2$p$-O at KO layer. Thus, in the 2DEG at the oxyfluoride superlattices, we have highly mobile carriers lighter than the ones reported in SrTiO$_3$/LaAlO$_3$ \cite{PhysRevB.80.045425}. 

In order to better explain the emergence of the 2DEG (2DHG) at the TaO$_2$/KF (ZnF$_2$/KO) interface, we applied an electrostatic model equivalent to the charged plates approach described in Ref  \cite{PhysRevB.80.045425}. First, let us define the total polarization in each compound as the sum of the initial (\emph{i.e.}, in zero field, $P_i^0$) polarization and the induced one (\emph{i.e} $P$ = $\epsilon_0 \chi_i \mathcal{E}_i$) by the effect of an applied electric field as follows:
 
\begin{equation}\label{eq:1}
	 P_i = P_i^0 + \epsilon_0 \chi_i \mathcal{E}_i,
\end{equation}

where the index $i$ = 1 and 2 stands for KTO and KZF materials respectively. Here, $\epsilon_0$ is the vacuum dielectric permittivity, $\chi_i$, and $\mathcal{E}_i$ are the dielectric susceptibility and electric field for each material, respectively. Now, considering short-circuit boundary conditions (\emph{i.e.}, $L_1$$\mathcal{E}_1$ = --$L_2$$\mathcal{E}_2$, where $L_1$ and $L_2$ are the KTO and KZF thicknesses) and the continuity of the displacement field vector, $\mathcal{D}_i$, at the interface (\emph{i.e.}, $\epsilon_0$$\mathcal{E}_1$+$P_1$ = $\epsilon_0$$\mathcal{E}_2$+$P_2$) we have:

\begin{eqnarray}
 	\mathcal{E}_1 & = & -\frac{L_2 (P_1^0 - P_2^0)}{\epsilon_0 (\epsilon_1 L_2 + \epsilon_2 L_1)}, \label{eq:2} 
	\\[10pt]
	\mathcal{E}_2 & = & +\frac{L_1 (P_1^0 - P_2^0)}{\epsilon_0 (\epsilon_1 L_2 + \epsilon_2 L_1)}, \label{eq:3}
\end{eqnarray}

here, $\epsilon_1$ and $\epsilon_2$ are the dielectric constants for the KTO and KZF bulk compounds. Then, obtaining the polarization and dielectric constants of the materials at zero field and, at the bulk level, the electric field developed could be computed at the superlattice arrangement.
In the KZF case, we found that $P_2^0 = 0$. Meanwhile for KTO, in addition to the polarization induced by the charged layers (\emph{i.e.}, $e$/2$\cdot\Theta$, where $\Theta$ is the unit-cell area), under biaxial strain conditions, this compound is expected to exhibit a paraelectric-to-ferroelectric transition due to its incipient-ferroelectric character \cite{PhysRevLett.104.227601}. Thus, the initial polarization is $P_1^0$ = ($e/2$$\cdot$$\Theta$ + $P^{KTO}_s$). However, the direction of such polarization will be strongly influenced by the character and value at  a given strain. 
Therefore, we considered three different cases, as follows:

\emph{Case 1:} Paraelectric KTO. Here, the superlattice $xy$ strain is imposed on the KTO bulk material, but, the paraelectric phase is still considered. Here, the computed KTO's dielectric constant, along the $z$-axis, is  $\epsilon_1$ = 93.66 and $P^{KTO}_s$ = 0. By using the Eqs. \ref{eq:2} and \ref{eq:3} we computed an electric field value of $\mathcal{E}_1$ $\approx$ $\mathcal{E}_2$ = 54.29 mV$\cdot$\r{A}$^{-1}$

\emph{Case 2:} Ferroelectric KTO, Bulk material relaxed under $xy$-strain $\rightarrow$ $P_s$ = $P_{xy}$. Similar as in the previous case, but the $xy$ polarization is allowed to relaxed ($Amm2$ phase). Nevertheless, along the $z$-axis the $P^{KTO}_s$ = 0 and the KTO's dielectric constant is  $\epsilon_1$ = 128.34. Thus, we obtained that $\mathcal{E}_1$ $\approx$ $\mathcal{E}_2$ = 40.62 mV$\cdot$\r{A}$^{-1}$

\emph{Case 3:} Ferroelectric KTO, Bulk material under the superlattice volume $\rightarrow$ $P_s$ = $P_{z}$. Here, the lattice volume obtained in the superlattice, for the middle KTO, is fixed to the bulk calculations (\emph{i.e} $a$ = $b$ = 4.022 \r{A} and $c$ =  4.060 \r{A}). Here, the $c$-lattice is larger than the imposed $ab$-plane, and therefore, a spontaneous polarization is developed along the $z$-axis ($P4mm$ phase). We found that the polarization along $z$ is $P^{KTO}_s$ = 27.57 $\mu$C$\cdot$cm$^{-2}$ and  $\epsilon_1$ = 44.65 for which, $\mathcal{E}_1$ $\approx$ $\mathcal{E}_2$ = 45.86 mV$\cdot$\r{A}$^{-1}$.

For all of the previous cases, $\epsilon_2$ = 9.41, $L_1$ = 28.49 \r{A} $\approx$ $L_2$ = 28.69 \r{A}. Additionally, $P^{KTO}_s$ was obtained by means of the Berry-phase approach \cite{Vanderbilt2000147}. These values are close to half from those obtained from the electric field computed from the DOS and the $ab$-initio electrostatic potential. In the following, we explain the source of the disagreement. 

\begin{figure}[t!]
 \centering
 \includegraphics[width=8.6cm,keepaspectratio=true]{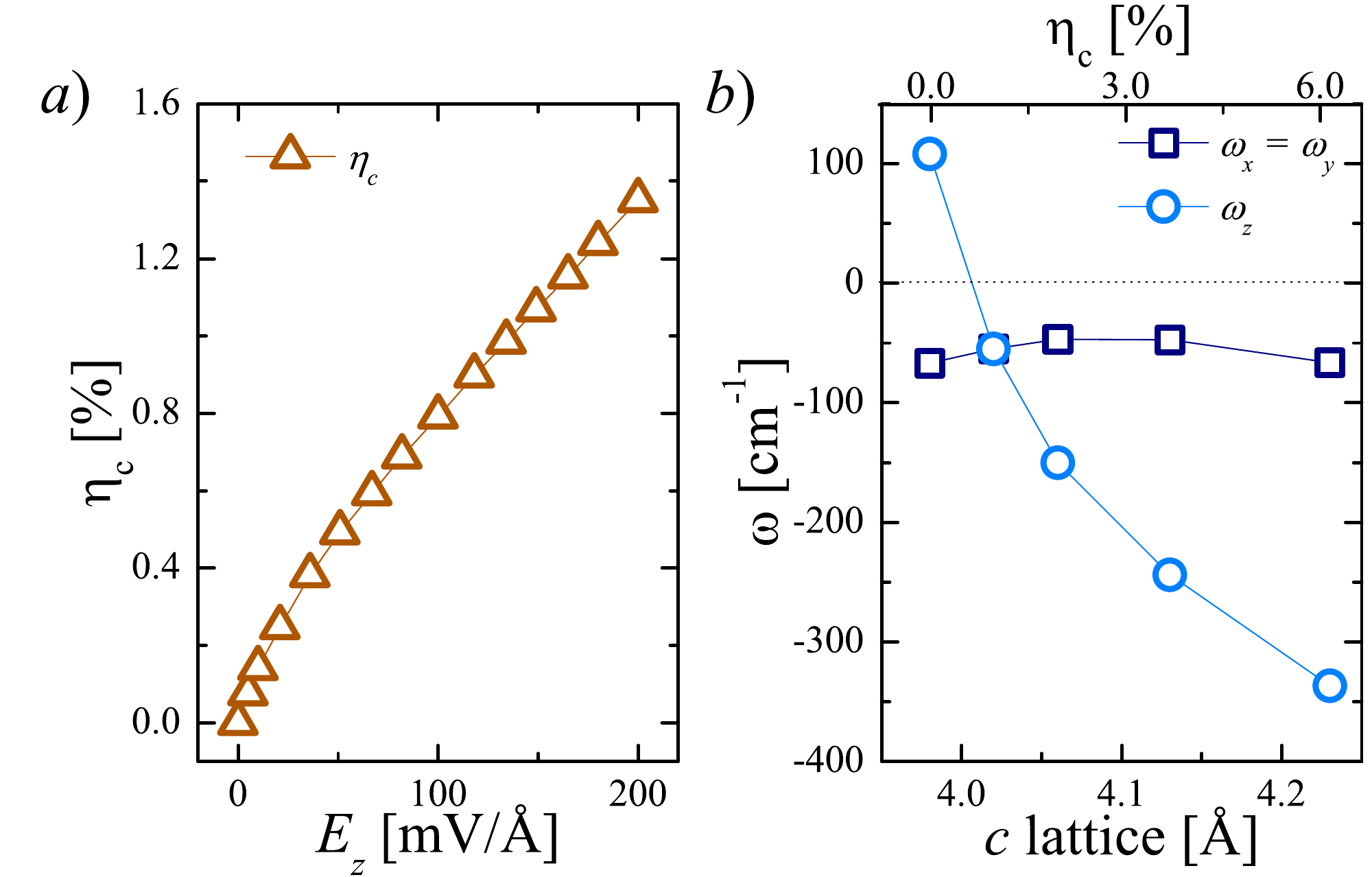}
 \caption{(Color online) $a$) Electric field induced elongation, $\eta_z$, along the $z$-axis at the KTaO$_3$. $b$) Unstable polar modes at $\Gamma$-point (shown at negative frequencies by notation), $\omega_x$ = $\omega_y$, and $\omega_z$, in the KTaO$_3$ as a function of the $c$-lattice elongation.}
 \label{fig:F2}
\end{figure} 

As it is well-known, and commented before, in bulk KTaO$_3$ the epitaxial strain can induce a stable ferroelectric state in agreement with its incipient-ferroelectric character. Nonetheless, the character of such state and polarization is determined by the nature of such strain. If the $xy$-plane biaxial strain is of expansion character, a 2-fold degenerated unstable polar phonon mode is observed at the $\Gamma$-point and an $Amm2$ ferroelectric ground state is achieved. In this phase, $P_s$ is pointing toward the [110] direction. 
When compressive epitaxial $xy$-strain is applied, a single polar phonon mode along the $z$-axis becomes unstable, which drives the system to a $P4mm$ ferroelectric ground state. Thus, under the $in$-plane expansion strain conditions in the KTO/K$M$F superlattices, an $Amm2$ ground state should be expected, and no traces of polarization should appear along the $z$-axis. Nevertheless and, as expected in dielectrics and ferroelectrics, the electric field might induces an expansion in the material that can renormalize its vibrational modes dynamics. 
Then, with the aim to explore how the internal electric field affects KTO, in Fig. \ref{fig:F2}$a$ we show the $c$-lattice parameter expansion as a function of the applied electric field in the same direction. This electric field is applied by keeping the $xy$-plane lattice parameters constant to the superlattice value and allowing only the $z$-axis and internal coordinates to relax. 
As one can observe, the $c$-lattice parameter experiences an expansion up to 1.4\% when the field is close to 200 mV$\cdot$\r{A}$^{-1}$ in agreement with the electrostriction phenomenon of the material.

On the second panel, Fig. \ref{fig:F2}$b$, we show the $\omega_{x}$=$\omega_{y}$ and $\omega_{z}$ unstable frequencies when expanding the $c$-lattice parameter of KTO. Interestingly, at the relaxed strain values, only the $xy$ modes are unstable suggesting the $Amm2$ phase in KTaO$_3$, as expected. However, when the $c$-lattice parameter is expanded (at larger values than the $in$-plane parameters, corresponding to strains above 0.90 \%), the polar-instability along the $z$-axis becomes stronger. This suggests that the $P4mm$ is the new ground state and therefore, switching of the polarization's direction occurs. 
The latter is the case of KTO, where, accordingly to our calculations, an expansion of $\approx$ 0.85\% can be obtained at an applied electric field $\mathcal{E}_z$ = 100 mV$\cdot$\r{A}$^{-1}$, as shown in Fig. \ref{fig:F2}$a$. At the oxyfluoride superlattice configuration, the KTO has a tetragonallity ratio of $c/a$ > 1, and therefore, the $P^{KTO}_s$ is flipped towards the $z$-axis. This observation could explain the disagreement between the results obtained from the computed electrostatic potential and its estimation through the electrostatic model when is applied to this system. Therefore, we conclude that the vibrational dynamics at the superlattices cannot be captured by the electrostatic model only based-on the bulk parameters. 

Such type of expansion has also been observed at the SrTiO$_3$/LaAlO$_3$ interface and it has been explained in terms of the electrostriction of the LaAlO$_3$ compound \cite{PhysRevLett.107.056102}.
Moreover, our reported behavior of the phonons, between bulk and superlattices, have been also highlighted in the BaTiO$_3$/BaO superlattices \cite{PhysRevB.82.045426}. 
Additionally, the Born effective charges, $Z^*$, of strained bulk KTO and at the KTO middle-layer of the 7/7 superlattice, are presented in Table \ref{tab:bec}. Here, it can be appreciated that the $Z^*_{Ta}$ along $z$ differs by almost 4 $e^-$ when compare the strained-bulk and the KTO at the superlattices. 
The latter finding is also suggesting a different dynamics between the bulk?s and superlattice?s eigendisplacements in relation to their phonon-modes.

\begin{table}[!t]
\caption{$zz$ components of the Born effective charges, $Z^*$ (in $e$) for the middle layers at the superlattices and components at the bulk compounds under strain.}
\centering
\begin{tabular}{c c c c c c c c c}
\hline
\hline
                   && $Z^*_K$  & $Z^*_M$ & $Z^*_X$$_\parallel$ & $Z^*_X$$_\bot$  \rule[-1ex]{0pt}{3.5ex}\\
\hline
Strained-bulk  &  KTaO$_3$ & +1.155 & +8.788  &  --6.633  & --1.655  \rule[-1ex]{0pt}{3.5ex}\\
                       &  KZnF$_3$ & +1.197 & +2.294  &  --1.698  & --0.896   \rule[-1ex]{0pt}{3.5ex}\\
\hline
Middle layer & KTaO$_3$ & +0.830  & +4.539  &  --4.009 & --0.953  \rule[-1ex]{0pt}{3.5ex}\\
                    & KZnF$_3$ & +1.548  & +2.967  &  --1.294 & --1.162\\
\hline
\hline
\end{tabular}
\label{tab:bec}
\end{table}

%%%%%%%%%%%%%%%%%%%%%%%%%
\subsection{Magnetic ordering at KTaO$_3$/KNiF$_3$ superlattices:}

We also analyzed the magnetic behavior of the oxyfluoride interfaces when $M$ = Ni. In bulk, KNiF$_3$ shows a $G$-type AFM ordering with a N\'eel temperature of 275 K \cite{doi:10.1143/JPSJ.14.1823}.Similarly, as it was done in the $M$ = Zn case, we studied the superlattices' thickness from 1/1 to 8/8 u.c. 

Initially, several magnetic orderings were tested, $G$-AFM, $C$-AFM, $A$-AFM, FM, and mixed $G$-AFM+FM layer (the FM configuration at the NiF$_2$ immediately after the TaO$_2$/KF interface and at the NiF$_2$/KO layers). According to our calculations, based-on hybrid-functionals performed with the \textsc{crystal} code, the ground-state magnetic ordering is always $G$-type AFM in the KNiF$_3$ film. 
These results are in full agreement with those obtained within the GGA+\emph{U} approximation for \emph{U} values above 6.0 eV with the \textsc{vasp} code. 
However, it is essential to mention that the mixed $G$-AFM+FM-layer state, (with the FM layer close to the NiF$_2$/KO interface), is very close in energy to the $G$-AFM ground state with $\Delta$E = 0.73 meV$\cdot$atom$^{-1}$. The $G$-AFM+FM magnetic state is even closer than other magnetic states such as the $A$-AFM and $C$-AFM.  
Moreover, we also found a weak magnetic moment of $m$ = 0.012 $\mu_B\cdot$atom$^{-1}$ localized on the Ta cations. These moments are following an FM interaction along $z$-axis with respect to the Ni sites ($m$ = 1.764 $\mu_B\cdot$atom$^{-1}$) in agreement with a Ni$^{+2}$ oxidation state. The latter is explained by the strongest FM interaction between half-occupied 3$d$--Ni:$e$\textsubscript{g} and the empty 5$d$--Ta:$t$\textsubscript{2g} \cite{Yu2012} rather than an AFM interaction driven by the super-exchange interaction between the Ni--O--Ta or Ta--F--Ni bonding.

Interestingly enough, there is a magnetic-moment modulation, of $\Delta$$m$ = 0.04 $\mu_B\cdot$atom$^{-1}$ between layers. This modulation observed across the entire thickness of the magnetic slab of the heterostructure in all the calculated superlattices.

%%%%%%%%%%%%%%%%%%%%%%%%%
\subsection{Cubic-Rashba spin-splitting at oxyfluoride interface:}

Recently, the appearance of spin-splitting followed by a complex spin-texture related to a Rashba-effect \cite{rashba1984} was reported to be attained at oxide interfaces such as SrTiO$_3$/LaAlO$_3$ \cite{PhysRevLett.104.126803, PhysRevLett.109.196401, PhysRevB.87.161102, Hurand2015}. Moreover, some reports have shown similar behavior in SrTiO$_3$-  and KTaO$_3$-based transistors \cite{PhysRevLett.108.206601, PhysRevB.80.121308}. 
In these single perovskite-based devices, a $k$-cubic dependence of the splitting was observed (see Eq. \ref{eq:6.1}) in contrast to the linear plus cubic dependence of the normal Rashba splitting occurring  at LaAlO$_3$/SrTiO$_3$ interface, and mainly due to the $d_{xy}$ -- $d_{xz/yz}$ multiorbital nature of the lowest bands \cite{PhysRevB.87.161102}. 
Additionally, in the case of Au(111) metal surface, it has been shown that even when the surface states exhibit mainly $p$--orbital character, the $d$-orbitals drives the Rashba splitting and dictates the direction of the spin by means of the orbital-angular momentum (OAM) \cite{PhysRevB.86.045437, Park20156}. 

\begin{figure*}[!t]
 \centering
 \includegraphics[width=14.0cm,keepaspectratio=true]{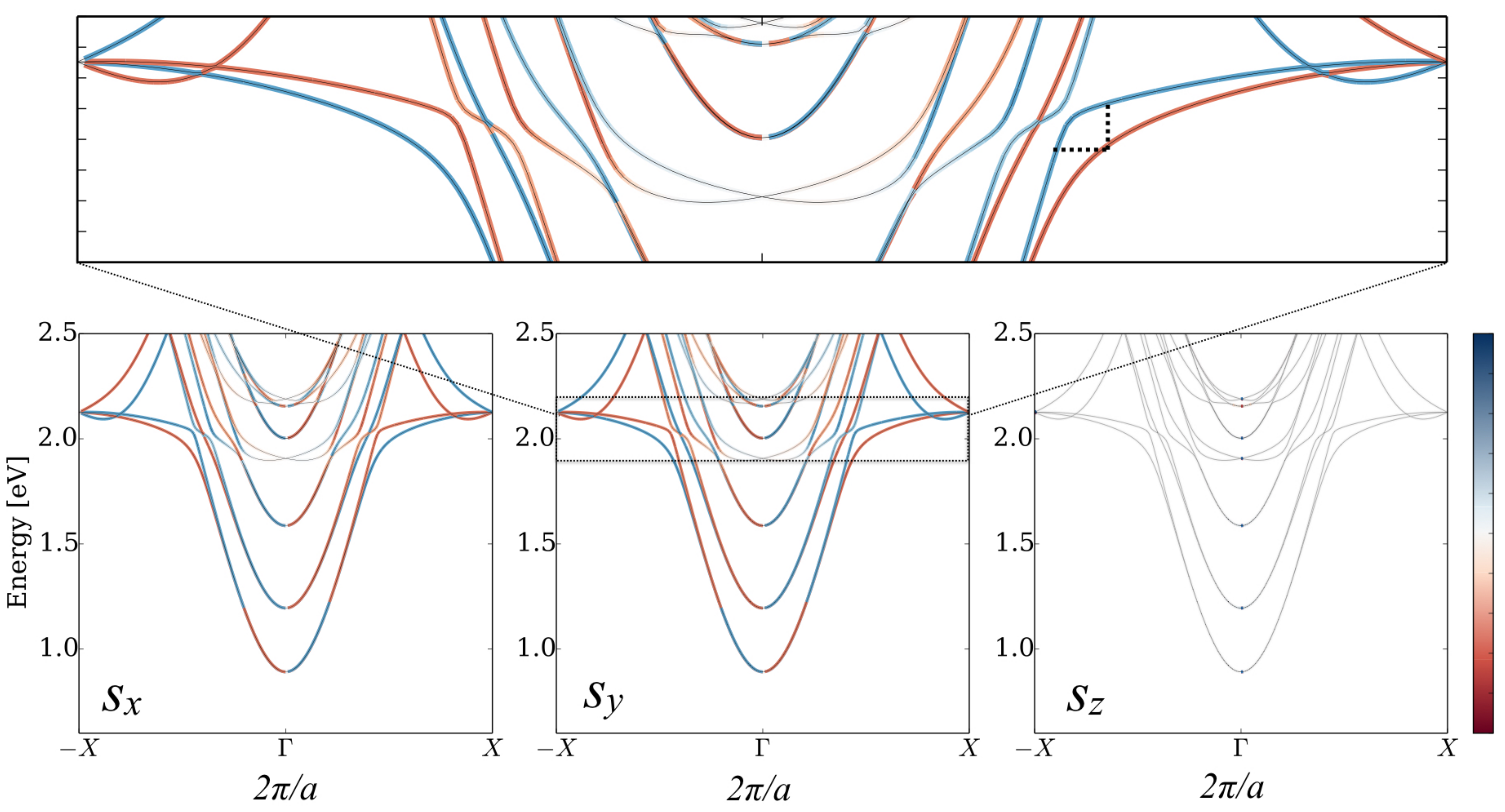}
 \caption{(Color online) Band structure for the $k_x$ path in which, $s_x$, $s_y$, and $s_z$ spin components are considered. The up and down orientations of the spin are represented by red and blue colors respectively. Figure obtained with the open-source code PyProcar \cite{herath2019pyprocar}.}
 \label{fig:rashba}
\end{figure*} 

Therefore, the transition metal $d$-orbitals at perovskite structures hold the key for a large and controllable electron band splitting with potential applications in future spintronic and spin-orbitronic devices \cite{PhysRevB.86.045437,Shanavas2014}. This latter property gives an additional degree of freedom to the already rich field of multifunctional perovskites. 

In order to look for the spin-texture and spin-orbit coupling (SOC) effects in the oxyfluoride interfaces, the analysis of the electronic structure is done by taking into account the SOC. Due to the large computational cost, the study of the bands was performed up to the 4/4 superlattice within the \textsc{vasp} code and analyzed by using the PyProcar package\cite{herath2019pyprocar}. %\textcolor{red}{E: With VASP I guess?}. 
These results can be traced up to the 7/7 system where, the 2DEG and 2DHG are formed, and the IMT takes place as shown previously. Thus, the results presented here can be extrapolated to larger $n$/$l$ superlattices. 

In Fig. \ref{fig:rashba} we can observe the electronic band structure along the $k_x$ path where the $s_x$, $s_y$, and $s_z$ spin components are presented separately and a color notation for the majority (red) and minority (blue) spin orientation is adopted. First, it can be noted that the electronic structure and the spin-texture at the conduction bands above of the Fermi level (and thus, close to the TaO$_2$/KF interface of the 7/7) are entirely 2D spin-polarized at the $k_x$--$k_y$ plane. Therefore, no band-dispersion is observed for the $s_z$ spin component. Surprisingly, a large spin-splitting of the bands is appreciated at the energy around 2.0 eV with respect to the Fermi energy. Additionally, the SOC effect is observed at the heavy $d_{xz/yz}$ crossing with the $d_{xy}$ lighter bands (see the upper inset in Fig. \ref{fig:rashba}).

The observed Rashba splitting is of the cubic-type character as observed in oxide surfaces, and interfaces and it is expressed by the Hamiltonian presented in Eq. \ref{eq:6.1}:

 \begin{equation}\label{eq:6.1}
	 H_{R3} = \alpha_3 \mathcal{E}_z i(k^3_{-} \sigma_{+} - k^3_{+}\sigma_{-}),
 \end{equation}

where $\alpha_3$ is the cubic-Rashba coupling term which measures the strength of the splitting (in units of eV$\cdot$$\r{A}^3$),  $\sigma_{\pm}$ = 1/2($\sigma_x$ $\pm$ $i$$\sigma_y$) and  $\sigma_{x,y}$ are the Pauli matrices. The $\mathcal{E}_z$ is the electric field perpendicular to the electron's plane movement denoted by $k_{\pm}$ = ($k_x$ $\pm$ $i$$k_y$) \cite{PhysRevLett.108.206601}. Here, the $E$ vs. $k$ profile for the bands is represented by the relationship:

 \begin{equation}\label{eq:6.2}
	 E^{\pm}(k) = \frac{\hbar^2 k^2}{2m^*}  \pm  \alpha_3 k^3.
 \end{equation} 
 
This type of spin-splitting can be understood like an opposite rotation of the lowest spin--up band with respect of the spin-down around an axis fixed at the vertices of the parabolic $d$-bands. 
This is in contrast with the linear horizontal displacement (or offset) of the spin--up and spin--down bands respectively observed in the most common linear-Rashba effect \footnote{The linear-Rashba effect has the $E$ vs. $k$ form: $E^{\pm}$($k$) = ($\hbar^2$$k^2$/$2m^*$) $\pm$ $\alpha$|$k$| coming from the Hamiltonian of the form $H_{R}$ = $\alpha$$\mathcal{E}_z$$i(k_{-} \sigma_{+} - k_{+}\sigma_{-})$} present in materials such as BiTeI \cite{Ishizaka2011}.

To obtain an idea of the order of magnitude of the spin-splitting, we compare it with other perovskite-based systems as follows. At the energy value of 2.0 eV we found that $\Delta k$ = 0.044 \r{A}$^{-1}$ which is more than twice of the one found at the KTaO$_3$-based transistor (\emph{i.e.}, Au/parylene-Al/KTaO$_3$) \cite{PhysRevLett.108.117602}. Besides, the vertical energy difference between the up- and down- spins is $\Delta E$ = 64 meV at the same bands as shown in Fig. \ref{fig:rashba}.
To compare with oxide/oxide interface, we computed the spin-texture of the SrTiO$_3$/LaAlO$_3$ system under the same conditions in a 4/4 superlattice (not shown here). We found that $\Delta k$ = 0.011 \r{A}$^{-1}$ and a $\Delta E$ = 11 meV. Then, this spin-splitting is four-times smaller in comparison to 0.044 \r{A}$^{-1}$ and 64 meV at the oxyfluoride interface. 

To go deeper into the analysis, we first computed the spontaneous polarization, which is directly related to the electric field $\mathcal{E}_z$ in Eq. \ref{eq:6.1}. The layer-by-layer polarization is reported in Table \ref{tab:pol-4-4}. It is observed that the polarization is in average close to $-$2.8 $\mu$C$\cdot$cm$^{-2}$ per KTO layer. For this case, $n$ = $l$ = 4, the total polarization is $-$11.14 $\mu$C$\cdot$cm$^{-2}$ at the oxide side. Moreover, the polarization is in the opposite direction of the fluoride layer with a value of $+$5.35 $\mu$C$\cdot$cm$^{-2}$. Then, at the end, a total polarization of $-$5.79 $\mu$C$\cdot$cm$^{-2}$ is still present along the entire superlattice structure. This significant difference suggests that the depolarizing field is not enough to compensate for the polarization along the entire supercell. Besides, the opposite directions of the polarizations is explained in the breaking of symmetry due to the O--M--F bonding along $z$.

Therefore, by increasing the number layers (\emph{i.e.}, the $n/l$ ratio), the KTaO$_3$ polarization will increase and then, it is expected that the spin-splitting will increase directly in agreement with Eq. \ref{eq:6.1}. 
Ultimately, it is important to note that the SOC in the Ta is at least 11 times larger than the one in the Ti cation. This can be taken as an advantage in the mixed oxide/fluoride interfaces due to the large lattice parameters of the fluorides perovskites that allows the incorporation of bigger perovskite oxides with 4$d$ and 5$d$ orbitals in novel superlattices and interfaces with mixed properties.

\begin{table}[!t]
\caption{Layer-by-layer spontaneous polarization along the $z$-axis ($P_z$ in $\mu$C$\cdot$cm$^{-2}$) in the 4/4 superlattice is shown. Besides, the oxide and fluoride polarizations are $P_{KTaO_3}$ = -11.14 $\mu$C$\cdot$cm$^{-2}$ and  $P_{KZnF_3}$ = 5.35 $\mu$C$\cdot$cm$^{-2}$ respectively.}
\centering
\begin{tabular}{c c c c c | c c c c}
\hline
\hline
 & \multicolumn{4}{c}{\emph{KTaO$_3$}} &  \multicolumn{4}{c}{\emph{KZnF$_3$}}  \rule[-1ex]{0pt}{3.5ex} \\
 \hline
Layer  & 1 &  2 & 3 & 4 & 5 & 6 & 7 & 8 \rule[-1ex]{0pt}{3.5ex} \\
\hline
$P_z$ & -2.85  & -2.52  &  -2.72  &  -2.86   & 1.43  &  1.15  &  1.20  &  1.58   \rule[-1ex]{0pt}{3.5ex}\\
\hline
\hline
\end{tabular}
\label{tab:pol-4-4}
\end{table}

From the magnetic KTaO$_3$/KNiF$_3$  superlattice, we noticed that the same spin-texture remains at the TaO$_2$/KO. Nonetheless, the spin-texture is becoming more complicated due to the overlapping of the 5$d$:Ta band with the unoccupied 3$d$($e$\textsubscript{g}):Ni states above the Fermi level.
It is essential to take advantage of the fact that the magnetic properties can be conserved in KNiF$_3$. Then, mixed multifunctional properties of 2DEG+2DHG+$k^3$-Rashba-splitting+$G$-AFM could be condensed in the same heterostructure, and then, other possible properties can be engineered in such systems which worth to be investigated in future works. 

Moreover, it worth noticing that the relevance of the $k^3$-Rashba dependence has been highlighted in the field of multifunctional properties based on the significant difference in the effective field symmetry between the $k$-linear and the $k^3$ Rashba. Thus, the $k^3$ symmetry influences all of the SOC related phenomena in materials, which, is not the case of the $k$-linear Rashba term. For example, in case of the spin Hall effect, it has been predicted that the $k$-cubic Rashba term can give rise to a more significant spin Hall conductivity \cite{PhysRevB.74.195330, PhysRevLett.113.086601}.

Finally, in this system, the possible tuning of the energy position of the 5$d$:Ta bands with respect to the Fermi energy, by controlling the $n/l$ ratio, stand as an additional feature of these systems. Therefore, the amount of splitting and consequently, the spin-transport properties can be tuned as well, which makes this type of interfaces even more appealing.\\

%%%%%%%%%%%%%%%%%%%%%%%%%
\section{Conclusions:}

We have analyzed employing the first-principles calculations the electronic and structural properties of oxyfluorides (KTaO$_3$)$_n$/(K$M$F$_3$)$_l$ $M$ = Zn and Ni interfaces.
We found that the polar catastrophe phenomenon takes place in these oxyfluoride interfaces. However, we found that the critical thickness for the appearance of 2DEG is obtained for a lower number of atomic layers than the LAO/STO oxide/oxide superlattices. The electron or hole gas is mostly constrained into the oxide part of the heterostructure.
We also observed that the magnetism in the KTaO$_3$/KNiF$_3$ exhibits a moment magnitude modulation but keeping on average the bulk $G$-type AFM.
Surprisingly, we observed a large $k^3$-Rashba type splitting at the oxyfluoride interfaces, at least four times larger than the one reported in SrTiO$_3$/LaAlO$_3$ interface and twice of the KTaO$_3$-based transistor. 
Thus, we believe that the oxyfluoride superlattices can be highlighted as novel perovskite-based systems in which, rich multifunctional properties can be tuned as in their oxides/oxides counterparts.\\

%%%%%%%%%%%%%%%%%%%%%%%%%
\emph{\textbf{Acknowledgements:}}

This work used XSEDE facilities which are supported by the National Science Foundation under grant number ACI-1053575. The authors also acknowledge the support from the Texas Advances Computer Center (with the Stampede2 and Bridges supercomputers),  the PRACE project TheDeNoMo and on the CECI facilities funded by F.R.S-FNRS (Grant No. 2.5020.1) and Tier-1 supercomputer of the F\'ed\'eration Wallonie-Bruxelles funded by the Walloon Region (Grant No. 1117545). This work was supported by the DMREF-NSF 1434897, NSF OAC-1740111 and DOE DE-SC0016176 projects. We acknowledge the West Virginia University supercomputing clusters; Spruce Knob and Thorny Flat which were used for the development of the library. 

%%%%%%%%%%%%%%%%%%%%%%%%%
\bibliography{library}

\begin{thebibliography}{74}%
\makeatletter
\providecommand \@ifxundefined [1]{%
 \@ifx{#1\undefined}
}%
\providecommand \@ifnum [1]{%
 \ifnum #1\expandafter \@firstoftwo
 \else \expandafter \@secondoftwo
 \fi
}%
\providecommand \@ifx [1]{%
 \ifx #1\expandafter \@firstoftwo
 \else \expandafter \@secondoftwo
 \fi
}%
\providecommand \natexlab [1]{#1}%
\providecommand \enquote  [1]{``#1''}%
\providecommand \bibnamefont  [1]{#1}%
\providecommand \bibfnamefont [1]{#1}%
\providecommand \citenamefont [1]{#1}%
\providecommand \href@noop [0]{\@secondoftwo}%
\providecommand \href [0]{\begingroup \@sanitize@url \@href}%
\providecommand \@href[1]{\@@startlink{#1}\@@href}%
\providecommand \@@href[1]{\endgroup#1\@@endlink}%
\providecommand \@sanitize@url [0]{\catcode `\\12\catcode `\$12\catcode
  `\&12\catcode `\#12\catcode `\^12\catcode `\_12\catcode `\%12\relax}%
\providecommand \@@startlink[1]{}%
\providecommand \@@endlink[0]{}%
\providecommand \url  [0]{\begingroup\@sanitize@url \@url }%
\providecommand \@url [1]{\endgroup\@href {#1}{\urlprefix }}%
\providecommand \urlprefix  [0]{URL }%
\providecommand \Eprint [0]{\href }%
\providecommand \doibase [0]{http://dx.doi.org/}%
\providecommand \selectlanguage [0]{\@gobble}%
\providecommand \bibinfo  [0]{\@secondoftwo}%
\providecommand \bibfield  [0]{\@secondoftwo}%
\providecommand \translation [1]{[#1]}%
\providecommand \BibitemOpen [0]{}%
\providecommand \bibitemStop [0]{}%
\providecommand \bibitemNoStop [0]{.\EOS\space}%
\providecommand \EOS [0]{\spacefactor3000\relax}%
\providecommand \BibitemShut  [1]{\csname bibitem#1\endcsname}%
\let\auto@bib@innerbib\@empty
%</preamble>
\bibitem [{\citenamefont {Hwang}\ \emph {et~al.}(2012)\citenamefont {Hwang},
  \citenamefont {Iwasa}, \citenamefont {Kawasaki}, \citenamefont {Keimer},
  \citenamefont {Nagaosa},\ and\ \citenamefont {Tokura}}]{Hwang2012}%
  \BibitemOpen
  \bibfield  {author} {\bibinfo {author} {\bibfnamefont {H.~Y.}\ \bibnamefont
  {Hwang}}, \bibinfo {author} {\bibfnamefont {Y.}~\bibnamefont {Iwasa}},
  \bibinfo {author} {\bibfnamefont {M.}~\bibnamefont {Kawasaki}}, \bibinfo
  {author} {\bibfnamefont {B.}~\bibnamefont {Keimer}}, \bibinfo {author}
  {\bibfnamefont {N.}~\bibnamefont {Nagaosa}}, \ and\ \bibinfo {author}
  {\bibfnamefont {Y.}~\bibnamefont {Tokura}},\ }\href {\doibase
  10.1038/nmat3223} {\bibfield  {journal} {\bibinfo  {journal} {Nature
  Materials}\ }\textbf {\bibinfo {volume} {11}},\ \bibinfo {pages} {103}
  (\bibinfo {year} {2012})}\BibitemShut {NoStop}%
\bibitem [{\citenamefont {Granozio}\ \emph {et~al.}(2013)\citenamefont
  {Granozio}, \citenamefont {Koster},\ and\ \citenamefont
  {Rijnders}}]{Granozio2013}%
  \BibitemOpen
  \bibfield  {author} {\bibinfo {author} {\bibfnamefont {F.}~\bibnamefont
  {Granozio}}, \bibinfo {author} {\bibfnamefont {G.}~\bibnamefont {Koster}}, \
  and\ \bibinfo {author} {\bibfnamefont {G.}~\bibnamefont {Rijnders}},\ }\href
  {http://journals.cambridge.org/abstract{\_}S0883769413002820} {\bibfield
  {journal} {\bibinfo  {journal} {MRS bulletin}\ }\textbf {\bibinfo {volume}
  {38}} (\bibinfo {year} {2013})}\BibitemShut {NoStop}%
\bibitem [{\citenamefont {Mannhart}\ \emph {et~al.}(2008)\citenamefont
  {Mannhart}, \citenamefont {Blank}, \citenamefont {Hwang}, \citenamefont
  {Millis},\ and\ \citenamefont {Triscone}}]{MRS:7956955}%
  \BibitemOpen
  \bibfield  {author} {\bibinfo {author} {\bibfnamefont {J.}~\bibnamefont
  {Mannhart}}, \bibinfo {author} {\bibfnamefont {D.}~\bibnamefont {Blank}},
  \bibinfo {author} {\bibfnamefont {H.}~\bibnamefont {Hwang}}, \bibinfo
  {author} {\bibfnamefont {A.}~\bibnamefont {Millis}}, \ and\ \bibinfo {author}
  {\bibfnamefont {J.-M.}\ \bibnamefont {Triscone}},\ }\href {\doibase
  10.1557/mrs2008.222} {\bibfield  {journal} {\bibinfo  {journal} {MRS
  Bulletin}\ }\textbf {\bibinfo {volume} {33}},\ \bibinfo {pages} {1027}
  (\bibinfo {year} {2008})}\BibitemShut {NoStop}%
\bibitem [{\citenamefont {Joshua}\ \emph {et~al.}(2012)\citenamefont {Joshua},
  \citenamefont {Pecker}, \citenamefont {Ruhman}, \citenamefont {Altman},\ and\
  \citenamefont {Ilani}}]{Joshua2012}%
  \BibitemOpen
  \bibfield  {author} {\bibinfo {author} {\bibfnamefont {A.}~\bibnamefont
  {Joshua}}, \bibinfo {author} {\bibfnamefont {S.}~\bibnamefont {Pecker}},
  \bibinfo {author} {\bibfnamefont {J.}~\bibnamefont {Ruhman}}, \bibinfo
  {author} {\bibfnamefont {E.}~\bibnamefont {Altman}}, \ and\ \bibinfo {author}
  {\bibfnamefont {S.}~\bibnamefont {Ilani}},\ }\href {\doibase
  10.1038/ncomms2116} {\bibfield  {journal} {\bibinfo  {journal} {Nature
  communications}\ }\textbf {\bibinfo {volume} {3}},\ \bibinfo {pages} {1129}
  (\bibinfo {year} {2012})}\BibitemShut {NoStop}%
\bibitem [{\citenamefont {Caprara}\ \emph {et~al.}(2012)\citenamefont
  {Caprara}, \citenamefont {Peronaci},\ and\ \citenamefont
  {Grilli}}]{PhysRevLett.109.196401}%
  \BibitemOpen
  \bibfield  {author} {\bibinfo {author} {\bibfnamefont {S.}~\bibnamefont
  {Caprara}}, \bibinfo {author} {\bibfnamefont {F.}~\bibnamefont {Peronaci}}, \
  and\ \bibinfo {author} {\bibfnamefont {M.}~\bibnamefont {Grilli}},\ }\href
  {\doibase 10.1103/PhysRevLett.109.196401} {\bibfield  {journal} {\bibinfo
  {journal} {Phys. Rev. Lett.}\ }\textbf {\bibinfo {volume} {109}},\ \bibinfo
  {pages} {196401} (\bibinfo {year} {2012})}\BibitemShut {NoStop}%
\bibitem [{\citenamefont {Zubko}\ \emph {et~al.}(2011)\citenamefont {Zubko},
  \citenamefont {Gariglio}, \citenamefont {Gabay}, \citenamefont {Ghosez},\
  and\ \citenamefont {Triscone}}]{Zubko2011}%
  \BibitemOpen
  \bibfield  {author} {\bibinfo {author} {\bibfnamefont {P.}~\bibnamefont
  {Zubko}}, \bibinfo {author} {\bibfnamefont {S.}~\bibnamefont {Gariglio}},
  \bibinfo {author} {\bibfnamefont {M.}~\bibnamefont {Gabay}}, \bibinfo
  {author} {\bibfnamefont {P.}~\bibnamefont {Ghosez}}, \ and\ \bibinfo {author}
  {\bibfnamefont {J.-M.}\ \bibnamefont {Triscone}},\ }\href {\doibase
  10.1146/annurev-conmatphys-062910-140445} {\bibfield  {journal} {\bibinfo
  {journal} {Annual Review of Condensed Matter Physics}\ }\textbf {\bibinfo
  {volume} {2}},\ \bibinfo {pages} {141} (\bibinfo {year} {2011})}\BibitemShut
  {NoStop}%
\bibitem [{\citenamefont {Nanda}\ and\ \citenamefont
  {Satpathy}(2010)}]{Nanda2010}%
  \BibitemOpen
  \bibfield  {author} {\bibinfo {author} {\bibfnamefont {B.~R.~K.}\
  \bibnamefont {Nanda}}\ and\ \bibinfo {author} {\bibfnamefont
  {S.}~\bibnamefont {Satpathy}},\ }\href {\doibase 10.1103/PhysRevB.81.224408}
  {\bibfield  {journal} {\bibinfo  {journal} {Phys. Rev. B}\ }\textbf {\bibinfo
  {volume} {81}},\ \bibinfo {pages} {224408} (\bibinfo {year}
  {2010})}\BibitemShut {NoStop}%
\bibitem [{\citenamefont {Bousquet}\ \emph {et~al.}(2008)\citenamefont
  {Bousquet}, \citenamefont {Dawber}, \citenamefont {Stucki}, \citenamefont
  {Lichtensteiger}, \citenamefont {Hermet}, \citenamefont {Gariglio},
  \citenamefont {Triscone},\ and\ \citenamefont {Ghosez}}]{Bousquet2008}%
  \BibitemOpen
  \bibfield  {author} {\bibinfo {author} {\bibfnamefont {E.}~\bibnamefont
  {Bousquet}}, \bibinfo {author} {\bibfnamefont {M.}~\bibnamefont {Dawber}},
  \bibinfo {author} {\bibfnamefont {N.}~\bibnamefont {Stucki}}, \bibinfo
  {author} {\bibfnamefont {C.}~\bibnamefont {Lichtensteiger}}, \bibinfo
  {author} {\bibfnamefont {P.}~\bibnamefont {Hermet}}, \bibinfo {author}
  {\bibfnamefont {S.}~\bibnamefont {Gariglio}}, \bibinfo {author}
  {\bibfnamefont {J.-M.}\ \bibnamefont {Triscone}}, \ and\ \bibinfo {author}
  {\bibfnamefont {P.}~\bibnamefont {Ghosez}},\ }\href {\doibase
  10.1038/nature06817} {\bibfield  {journal} {\bibinfo  {journal} {Nature}\
  }\textbf {\bibinfo {volume} {452}},\ \bibinfo {pages} {732} (\bibinfo {year}
  {2008})}\BibitemShut {NoStop}%
\bibitem [{\citenamefont {Benedek}\ and\ \citenamefont
  {Fennie}(2011)}]{Benedek2011}%
  \BibitemOpen
  \bibfield  {author} {\bibinfo {author} {\bibfnamefont {N.}~\bibnamefont
  {Benedek}}\ and\ \bibinfo {author} {\bibfnamefont {C.}~\bibnamefont
  {Fennie}},\ }\href {\doibase 10.1103/PhysRevLett.106.107204} {\bibfield
  {journal} {\bibinfo  {journal} {Phys. Rev. Lett.}\ }\textbf {\bibinfo
  {volume} {106}},\ \bibinfo {pages} {107204} (\bibinfo {year}
  {2011})}\BibitemShut {NoStop}%
\bibitem [{\citenamefont {Mulder}\ \emph {et~al.}(2013)\citenamefont {Mulder},
  \citenamefont {Benedek}, \citenamefont {Rondinelli},\ and\ \citenamefont
  {Fennie}}]{ADFM:ADFM201300210}%
  \BibitemOpen
  \bibfield  {author} {\bibinfo {author} {\bibfnamefont {A.~T.}\ \bibnamefont
  {Mulder}}, \bibinfo {author} {\bibfnamefont {N.~A.}\ \bibnamefont {Benedek}},
  \bibinfo {author} {\bibfnamefont {J.~M.}\ \bibnamefont {Rondinelli}}, \ and\
  \bibinfo {author} {\bibfnamefont {C.~J.}\ \bibnamefont {Fennie}},\ }\href
  {\doibase 10.1002/adfm.201300210} {\bibfield  {journal} {\bibinfo  {journal}
  {Advanced Functional Materials}\ }\textbf {\bibinfo {volume} {23}},\ \bibinfo
  {pages} {4810} (\bibinfo {year} {2013})}\BibitemShut {NoStop}%
\bibitem [{\citenamefont {Zanolli}\ \emph {et~al.}(2013)\citenamefont
  {Zanolli}, \citenamefont {Wojdeł}, \citenamefont {\'{I}\~{n}iguez},\ and\
  \citenamefont {Ghosez}}]{Zanolli2013}%
  \BibitemOpen
  \bibfield  {author} {\bibinfo {author} {\bibfnamefont {Z.}~\bibnamefont
  {Zanolli}}, \bibinfo {author} {\bibfnamefont {J.~C.}\ \bibnamefont
  {Wojdeł}}, \bibinfo {author} {\bibfnamefont {J.}~\bibnamefont
  {\'{I}\~{n}iguez}}, \ and\ \bibinfo {author} {\bibfnamefont {P.}~\bibnamefont
  {Ghosez}},\ }\href {\doibase 10.1103/PhysRevB.88.060102} {\bibfield
  {journal} {\bibinfo  {journal} {Phys. Rev. B}\ }\textbf {\bibinfo {volume}
  {88}},\ \bibinfo {pages} {060102} (\bibinfo {year} {2013})}\BibitemShut
  {NoStop}%
\bibitem [{\citenamefont {Garcia-Castro}\ \emph {et~al.}(2014)\citenamefont
  {Garcia-Castro}, \citenamefont {Spaldin}, \citenamefont {Romero},\ and\
  \citenamefont {Bousquet}}]{PhysRevB.89.104107}%
  \BibitemOpen
  \bibfield  {author} {\bibinfo {author} {\bibfnamefont {A.~C.}\ \bibnamefont
  {Garcia-Castro}}, \bibinfo {author} {\bibfnamefont {N.~A.}\ \bibnamefont
  {Spaldin}}, \bibinfo {author} {\bibfnamefont {A.~H.}\ \bibnamefont {Romero}},
  \ and\ \bibinfo {author} {\bibfnamefont {E.}~\bibnamefont {Bousquet}},\
  }\href {\doibase 10.1103/PhysRevB.89.104107} {\bibfield  {journal} {\bibinfo
  {journal} {Phys. Rev. B}\ }\textbf {\bibinfo {volume} {89}},\ \bibinfo
  {pages} {104107} (\bibinfo {year} {2014})}\BibitemShut {NoStop}%
\bibitem [{\citenamefont {Garcia-Castro}\ \emph {et~al.}(2016)\citenamefont
  {Garcia-Castro}, \citenamefont {Romero},\ and\ \citenamefont
  {Bousquet}}]{PhysRevLett.116.117202}%
  \BibitemOpen
  \bibfield  {author} {\bibinfo {author} {\bibfnamefont {A.~C.}\ \bibnamefont
  {Garcia-Castro}}, \bibinfo {author} {\bibfnamefont {A.~H.}\ \bibnamefont
  {Romero}}, \ and\ \bibinfo {author} {\bibfnamefont {E.}~\bibnamefont
  {Bousquet}},\ }\href {\doibase 10.1103/PhysRevLett.116.117202} {\bibfield
  {journal} {\bibinfo  {journal} {Phys. Rev. Lett.}\ }\textbf {\bibinfo
  {volume} {116}},\ \bibinfo {pages} {117202} (\bibinfo {year}
  {2016})}\BibitemShut {NoStop}%
\bibitem [{\citenamefont {Yang}\ \emph {et~al.}(2017)\citenamefont {Yang},
  \citenamefont {KC}, \citenamefont {Garcia-Castro}, \citenamefont {Borisov},
  \citenamefont {Bousquet}, \citenamefont {Lederman}, \citenamefont {Romero},\
  and\ \citenamefont {Cen}}]{Yang2017}%
  \BibitemOpen
  \bibfield  {author} {\bibinfo {author} {\bibfnamefont {M.}~\bibnamefont
  {Yang}}, \bibinfo {author} {\bibfnamefont {A.}~\bibnamefont {KC}}, \bibinfo
  {author} {\bibfnamefont {A.~C.}\ \bibnamefont {Garcia-Castro}}, \bibinfo
  {author} {\bibfnamefont {P.}~\bibnamefont {Borisov}}, \bibinfo {author}
  {\bibfnamefont {E.}~\bibnamefont {Bousquet}}, \bibinfo {author}
  {\bibfnamefont {D.}~\bibnamefont {Lederman}}, \bibinfo {author}
  {\bibfnamefont {A.~H.}\ \bibnamefont {Romero}}, \ and\ \bibinfo {author}
  {\bibfnamefont {C.}~\bibnamefont {Cen}},\ }\href {\doibase
  10.1038/s41598-017-07834-0} {\bibfield  {journal} {\bibinfo  {journal}
  {Scientific Reports}\ }\textbf {\bibinfo {volume} {7}},\ \bibinfo {pages}
  {7182} (\bibinfo {year} {2017})}\BibitemShut {NoStop}%
\bibitem [{\citenamefont {Carpenter}\ \emph {et~al.}(2012)\citenamefont
  {Carpenter}, \citenamefont {Salje},\ and\ \citenamefont
  {Howard}}]{PhysRevB.85.224430}%
  \BibitemOpen
  \bibfield  {author} {\bibinfo {author} {\bibfnamefont {M.~A.}\ \bibnamefont
  {Carpenter}}, \bibinfo {author} {\bibfnamefont {E.~K.~H.}\ \bibnamefont
  {Salje}}, \ and\ \bibinfo {author} {\bibfnamefont {C.~J.}\ \bibnamefont
  {Howard}},\ }\href {\doibase 10.1103/PhysRevB.85.224430} {\bibfield
  {journal} {\bibinfo  {journal} {Phys. Rev. B}\ }\textbf {\bibinfo {volume}
  {85}},\ \bibinfo {pages} {224430} (\bibinfo {year} {2012})}\BibitemShut
  {NoStop}%
\bibitem [{\citenamefont {Dobson}\ \emph {et~al.}(2011)\citenamefont {Dobson},
  \citenamefont {Hunt}, \citenamefont {Lindsay-Scott},\ and\ \citenamefont
  {Wood}}]{Dobson2011171}%
  \BibitemOpen
  \bibfield  {author} {\bibinfo {author} {\bibfnamefont {D.~P.}\ \bibnamefont
  {Dobson}}, \bibinfo {author} {\bibfnamefont {S.~A.}\ \bibnamefont {Hunt}},
  \bibinfo {author} {\bibfnamefont {A.}~\bibnamefont {Lindsay-Scott}}, \ and\
  \bibinfo {author} {\bibfnamefont {I.~G.}\ \bibnamefont {Wood}},\ }\href
  {\doibase 10.1016/j.pepi.2011.08.010} {\bibfield  {journal} {\bibinfo
  {journal} {Physics of the Earth and Planetary Interiors}\ }\textbf {\bibinfo
  {volume} {189}},\ \bibinfo {pages} {171 } (\bibinfo {year}
  {2011})}\BibitemShut {NoStop}%
\bibitem [{\citenamefont {Edwardson}\ \emph {et~al.}(1989)\citenamefont
  {Edwardson}, \citenamefont {Boyer}, \citenamefont {Newman}, \citenamefont
  {Fox}, \citenamefont {Hardy}, \citenamefont {Flocken}, \citenamefont
  {Guenther},\ and\ \citenamefont {Mei}}]{PhysRevB.39.9738}%
  \BibitemOpen
  \bibfield  {author} {\bibinfo {author} {\bibfnamefont {P.~J.}\ \bibnamefont
  {Edwardson}}, \bibinfo {author} {\bibfnamefont {L.~L.}\ \bibnamefont
  {Boyer}}, \bibinfo {author} {\bibfnamefont {R.~L.}\ \bibnamefont {Newman}},
  \bibinfo {author} {\bibfnamefont {D.~H.}\ \bibnamefont {Fox}}, \bibinfo
  {author} {\bibfnamefont {J.~R.}\ \bibnamefont {Hardy}}, \bibinfo {author}
  {\bibfnamefont {J.~W.}\ \bibnamefont {Flocken}}, \bibinfo {author}
  {\bibfnamefont {R.~A.}\ \bibnamefont {Guenther}}, \ and\ \bibinfo {author}
  {\bibfnamefont {W.}~\bibnamefont {Mei}},\ }\href {\doibase
  10.1103/PhysRevB.39.9738} {\bibfield  {journal} {\bibinfo  {journal} {Phys.
  Rev. B}\ }\textbf {\bibinfo {volume} {39}},\ \bibinfo {pages} {9738}
  (\bibinfo {year} {1989})}\BibitemShut {NoStop}%
\bibitem [{\citenamefont {Duan}\ \emph {et~al.}(2004)\citenamefont {Duan},
  \citenamefont {Mei}, \citenamefont {Liu}, \citenamefont {Yin}, \citenamefont
  {Hardy}, \citenamefont {Smith}, \citenamefont {Mehl},\ and\ \citenamefont
  {Boyer}}]{PhysRevB.69.033102}%
  \BibitemOpen
  \bibfield  {author} {\bibinfo {author} {\bibfnamefont {C.-g.}\ \bibnamefont
  {Duan}}, \bibinfo {author} {\bibfnamefont {W.~N.}\ \bibnamefont {Mei}},
  \bibinfo {author} {\bibfnamefont {J.}~\bibnamefont {Liu}}, \bibinfo {author}
  {\bibfnamefont {W.-G.}\ \bibnamefont {Yin}}, \bibinfo {author} {\bibfnamefont
  {J.~R.}\ \bibnamefont {Hardy}}, \bibinfo {author} {\bibfnamefont {R.~W.}\
  \bibnamefont {Smith}}, \bibinfo {author} {\bibfnamefont {M.~J.}\ \bibnamefont
  {Mehl}}, \ and\ \bibinfo {author} {\bibfnamefont {L.~L.}\ \bibnamefont
  {Boyer}},\ }\href {\doibase 10.1103/PhysRevB.69.033102} {\bibfield  {journal}
  {\bibinfo  {journal} {Phys. Rev. B}\ }\textbf {\bibinfo {volume} {69}},\
  \bibinfo {pages} {033102} (\bibinfo {year} {2004})}\BibitemShut {NoStop}%
\bibitem [{\citenamefont {Pang}\ \emph {et~al.}(2004)\citenamefont {Pang},
  \citenamefont {Stamps}, \citenamefont {Malkinski}, \citenamefont {Celinski},\
  and\ \citenamefont {Skrzypek}}]{Pang2004}%
  \BibitemOpen
  \bibfield  {author} {\bibinfo {author} {\bibfnamefont {W.}~\bibnamefont
  {Pang}}, \bibinfo {author} {\bibfnamefont {R.~L.}\ \bibnamefont {Stamps}},
  \bibinfo {author} {\bibfnamefont {L.}~\bibnamefont {Malkinski}}, \bibinfo
  {author} {\bibfnamefont {Z.}~\bibnamefont {Celinski}}, \ and\ \bibinfo
  {author} {\bibfnamefont {D.}~\bibnamefont {Skrzypek}},\ }\href {\doibase
  http://dx.doi.org/10.1063/1.1687552} {\bibfield  {journal} {\bibinfo
  {journal} {Journal of Applied Physics}\ }\textbf {\bibinfo {volume} {95}},\
  \bibinfo {pages} {7309} (\bibinfo {year} {2004})}\BibitemShut {NoStop}%
\bibitem [{\citenamefont {Widuch}\ \emph {et~al.}(2011)\citenamefont {Widuch},
  \citenamefont {Stamps}, \citenamefont {Skrzypek},\ and\ \citenamefont
  {Celinski}}]{Widuch2011}%
  \BibitemOpen
  \bibfield  {author} {\bibinfo {author} {\bibfnamefont {S.}~\bibnamefont
  {Widuch}}, \bibinfo {author} {\bibfnamefont {R.~L.}\ \bibnamefont {Stamps}},
  \bibinfo {author} {\bibfnamefont {D.}~\bibnamefont {Skrzypek}}, \ and\
  \bibinfo {author} {\bibfnamefont {Z.}~\bibnamefont {Celinski}},\ }\href
  {\doibase 10.1088/0022-3727/44/41/415003} {\bibfield  {journal} {\bibinfo
  {journal} {Journal of Physics D: Applied Physics}\ }\textbf {\bibinfo
  {volume} {44}},\ \bibinfo {pages} {415003} (\bibinfo {year}
  {2011})}\BibitemShut {NoStop}%
\bibitem [{\citenamefont {Moon}\ \emph {et~al.}(2014)\citenamefont {Moon},
  \citenamefont {Xie}, \citenamefont {Laird}, \citenamefont {Keavney},
  \citenamefont {Li},\ and\ \citenamefont {May}}]{doi:10.1021/ja410954z}%
  \BibitemOpen
  \bibfield  {author} {\bibinfo {author} {\bibfnamefont {E.~J.}\ \bibnamefont
  {Moon}}, \bibinfo {author} {\bibfnamefont {Y.}~\bibnamefont {Xie}}, \bibinfo
  {author} {\bibfnamefont {E.~D.}\ \bibnamefont {Laird}}, \bibinfo {author}
  {\bibfnamefont {D.~J.}\ \bibnamefont {Keavney}}, \bibinfo {author}
  {\bibfnamefont {C.~Y.}\ \bibnamefont {Li}}, \ and\ \bibinfo {author}
  {\bibfnamefont {S.~J.}\ \bibnamefont {May}},\ }\href {\doibase
  10.1021/ja410954z} {\bibfield  {journal} {\bibinfo  {journal} {Journal of the
  American Chemical Society}\ }\textbf {\bibinfo {volume} {136}},\ \bibinfo
  {pages} {2224} (\bibinfo {year} {2014})}\BibitemShut {NoStop}%
\bibitem [{\citenamefont {AI-Mamouri}\ \emph {et~al.}(1994)\citenamefont
  {AI-Mamouri}, \citenamefont {Edwards}, \citenamefont {Greaves},\ and\
  \citenamefont {Slaski}}]{AI-Mamouri1994}%
  \BibitemOpen
  \bibfield  {author} {\bibinfo {author} {\bibfnamefont {M.}~\bibnamefont
  {AI-Mamouri}}, \bibinfo {author} {\bibfnamefont {P.~P.}\ \bibnamefont
  {Edwards}}, \bibinfo {author} {\bibfnamefont {C.}~\bibnamefont {Greaves}}, \
  and\ \bibinfo {author} {\bibfnamefont {M.}~\bibnamefont {Slaski}},\ }\href
  {\doibase 10.1038/369382a0} {\bibfield  {journal} {\bibinfo  {journal}
  {Nature}\ }\textbf {\bibinfo {volume} {369}},\ \bibinfo {pages} {382}
  (\bibinfo {year} {1994})}\BibitemShut {NoStop}%
\bibitem [{\citenamefont {Sturza}\ \emph {et~al.}(2010)\citenamefont {Sturza},
  \citenamefont {Daviero-Minaud}, \citenamefont {Kabbour}, \citenamefont
  {Gardoll},\ and\ \citenamefont {Mentre}}]{doi:10.1021/cm102724k}%
  \BibitemOpen
  \bibfield  {author} {\bibinfo {author} {\bibfnamefont {M.}~\bibnamefont
  {Sturza}}, \bibinfo {author} {\bibfnamefont {S.}~\bibnamefont
  {Daviero-Minaud}}, \bibinfo {author} {\bibfnamefont {H.}~\bibnamefont
  {Kabbour}}, \bibinfo {author} {\bibfnamefont {O.}~\bibnamefont {Gardoll}}, \
  and\ \bibinfo {author} {\bibfnamefont {O.}~\bibnamefont {Mentre}},\ }\href
  {\doibase 10.1021/cm102724k} {\bibfield  {journal} {\bibinfo  {journal}
  {Chemistry of Materials}\ }\textbf {\bibinfo {volume} {22}},\ \bibinfo
  {pages} {6726} (\bibinfo {year} {2010})}\BibitemShut {NoStop}%
\bibitem [{\citenamefont {Sturza}\ \emph {et~al.}(2011)\citenamefont {Sturza},
  \citenamefont {Kabbour}, \citenamefont {Daviero-Minaud}, \citenamefont
  {Filimonov}, \citenamefont {Pokholok}, \citenamefont {Tiercelin},
  \citenamefont {Porcher}, \citenamefont {Aldon},\ and\ \citenamefont
  {Mentre}}]{doi:10.1021/ja2028603}%
  \BibitemOpen
  \bibfield  {author} {\bibinfo {author} {\bibfnamefont {M.}~\bibnamefont
  {Sturza}}, \bibinfo {author} {\bibfnamefont {H.}~\bibnamefont {Kabbour}},
  \bibinfo {author} {\bibfnamefont {S.}~\bibnamefont {Daviero-Minaud}},
  \bibinfo {author} {\bibfnamefont {D.}~\bibnamefont {Filimonov}}, \bibinfo
  {author} {\bibfnamefont {K.}~\bibnamefont {Pokholok}}, \bibinfo {author}
  {\bibfnamefont {N.}~\bibnamefont {Tiercelin}}, \bibinfo {author}
  {\bibfnamefont {F.}~\bibnamefont {Porcher}}, \bibinfo {author} {\bibfnamefont
  {L.}~\bibnamefont {Aldon}}, \ and\ \bibinfo {author} {\bibfnamefont
  {O.}~\bibnamefont {Mentre}},\ }\href {\doibase 10.1021/ja2028603} {\bibfield
  {journal} {\bibinfo  {journal} {Journal of the American Chemical Society}\
  }\textbf {\bibinfo {volume} {133}},\ \bibinfo {pages} {10901} (\bibinfo
  {year} {2011})}\BibitemShut {NoStop}%
\bibitem [{\citenamefont {Yang}\ and\ \citenamefont {Su}(2014)}]{Yang2014}%
  \BibitemOpen
  \bibfield  {author} {\bibinfo {author} {\bibfnamefont {X.}~\bibnamefont
  {Yang}}\ and\ \bibinfo {author} {\bibfnamefont {H.}~\bibnamefont {Su}},\
  }\href {\doibase 10.1038/srep05420} {\bibfield  {journal} {\bibinfo
  {journal} {Scientific Reports}\ }\textbf {\bibinfo {volume} {4}},\ \bibinfo
  {pages} {5420} (\bibinfo {year} {2014})}\BibitemShut {NoStop}%
\bibitem [{\citenamefont {Yang}\ and\ \citenamefont {Su}(2015)}]{Yang2015}%
  \BibitemOpen
  \bibfield  {author} {\bibinfo {author} {\bibfnamefont {X.}~\bibnamefont
  {Yang}}\ and\ \bibinfo {author} {\bibfnamefont {H.}~\bibnamefont {Su}},\
  }\href {\doibase 10.1038/srep15849} {\bibfield  {journal} {\bibinfo
  {journal} {Scientific Reports}\ }\textbf {\bibinfo {volume} {5}},\ \bibinfo
  {pages} {15849} (\bibinfo {year} {2015})}\BibitemShut {NoStop}%
\bibitem [{\citenamefont {Deng}(2017)}]{CNMA:CNMA201600342}%
  \BibitemOpen
  \bibfield  {author} {\bibinfo {author} {\bibfnamefont {D.}~\bibnamefont
  {Deng}},\ }\href {\doibase 10.1002/cnma.201600342} {\bibfield  {journal}
  {\bibinfo  {journal} {ChemNanoMat}\ }\textbf {\bibinfo {volume} {3}},\
  \bibinfo {pages} {146} (\bibinfo {year} {2017})}\BibitemShut {NoStop}%
\bibitem [{\citenamefont {Charles}\ \emph {et~al.}(2018)\citenamefont
  {Charles}, \citenamefont {Saballos},\ and\ \citenamefont
  {Rondinelli}}]{Charles2018}%
  \BibitemOpen
  \bibfield  {author} {\bibinfo {author} {\bibfnamefont {N.}~\bibnamefont
  {Charles}}, \bibinfo {author} {\bibfnamefont {R.~J.}\ \bibnamefont
  {Saballos}}, \ and\ \bibinfo {author} {\bibfnamefont {J.~M.}\ \bibnamefont
  {Rondinelli}},\ }\href {\doibase 10.1021/acs.chemmater.8b01336} {\bibfield
  {journal} {\bibinfo  {journal} {Chemistry of Materials}\ }\textbf {\bibinfo
  {volume} {30}},\ \bibinfo {pages} {3528} (\bibinfo {year}
  {2018})}\BibitemShut {NoStop}%
\bibitem [{\citenamefont {Kageyama}\ \emph {et~al.}(2018)\citenamefont
  {Kageyama}, \citenamefont {Hayashi}, \citenamefont {Maeda}, \citenamefont
  {Attfield}, \citenamefont {Hiroi}, \citenamefont {Rondinelli},\ and\
  \citenamefont {Poeppelmeier}}]{Kageyama2018}%
  \BibitemOpen
  \bibfield  {author} {\bibinfo {author} {\bibfnamefont {H.}~\bibnamefont
  {Kageyama}}, \bibinfo {author} {\bibfnamefont {K.}~\bibnamefont {Hayashi}},
  \bibinfo {author} {\bibfnamefont {K.}~\bibnamefont {Maeda}}, \bibinfo
  {author} {\bibfnamefont {J.~P.}\ \bibnamefont {Attfield}}, \bibinfo {author}
  {\bibfnamefont {Z.}~\bibnamefont {Hiroi}}, \bibinfo {author} {\bibfnamefont
  {J.~M.}\ \bibnamefont {Rondinelli}}, \ and\ \bibinfo {author} {\bibfnamefont
  {K.~R.}\ \bibnamefont {Poeppelmeier}},\ }\href {\doibase
  10.1038/s41467-018-02838-4} {\bibfield  {journal} {\bibinfo  {journal}
  {Nature Communications}\ }\textbf {\bibinfo {volume} {9}} (\bibinfo {year}
  {2018}),\ 10.1038/s41467-018-02838-4}\BibitemShut {NoStop}%
\bibitem [{\citenamefont {Harada}\ \emph {et~al.}(2019)\citenamefont {Harada},
  \citenamefont {Charles}, \citenamefont {Poeppelmeier},\ and\ \citenamefont
  {Rondinelli}}]{Harada2019}%
  \BibitemOpen
  \bibfield  {author} {\bibinfo {author} {\bibfnamefont {J.~K.}\ \bibnamefont
  {Harada}}, \bibinfo {author} {\bibfnamefont {N.}~\bibnamefont {Charles}},
  \bibinfo {author} {\bibfnamefont {K.~R.}\ \bibnamefont {Poeppelmeier}}, \
  and\ \bibinfo {author} {\bibfnamefont {J.~M.}\ \bibnamefont {Rondinelli}},\
  }\href {\doibase 10.1002/adma.201805295} {\bibfield  {journal} {\bibinfo
  {journal} {Advanced Materials}\ }\textbf {\bibinfo {volume} {1805295}},\
  \bibinfo {pages} {1805295} (\bibinfo {year} {2019})}\BibitemShut {NoStop}%
\bibitem [{\citenamefont {Hohenberg}\ and\ \citenamefont
  {Kohn}(1964)}]{PhysRev.136.B864}%
  \BibitemOpen
  \bibfield  {author} {\bibinfo {author} {\bibfnamefont {P.}~\bibnamefont
  {Hohenberg}}\ and\ \bibinfo {author} {\bibfnamefont {W.}~\bibnamefont
  {Kohn}},\ }\href {\doibase 10.1103/PhysRev.136.B864} {\bibfield  {journal}
  {\bibinfo  {journal} {Phys. Rev.}\ }\textbf {\bibinfo {volume} {136}},\
  \bibinfo {pages} {B864} (\bibinfo {year} {1964})}\BibitemShut {NoStop}%
\bibitem [{\citenamefont {Kohn}\ and\ \citenamefont
  {Sham}(1965)}]{PhysRev.140.A1133}%
  \BibitemOpen
  \bibfield  {author} {\bibinfo {author} {\bibfnamefont {W.}~\bibnamefont
  {Kohn}}\ and\ \bibinfo {author} {\bibfnamefont {L.~J.}\ \bibnamefont
  {Sham}},\ }\href {\doibase 10.1103/PhysRev.140.A1133} {\bibfield  {journal}
  {\bibinfo  {journal} {Phys. Rev.}\ }\textbf {\bibinfo {volume} {140}},\
  \bibinfo {pages} {A1133} (\bibinfo {year} {1965})}\BibitemShut {NoStop}%
\bibitem [{\citenamefont {Kresse}\ and\ \citenamefont
  {Furthm\"{u}ller}(1996)}]{Kresse1996}%
  \BibitemOpen
  \bibfield  {author} {\bibinfo {author} {\bibfnamefont {G.}~\bibnamefont
  {Kresse}}\ and\ \bibinfo {author} {\bibfnamefont {J.}~\bibnamefont
  {Furthm\"{u}ller}},\ }\href {http://www.ncbi.nlm.nih.gov/pubmed/9984901}
  {\bibfield  {journal} {\bibinfo  {journal} {Phys. Rev. B}\ }\textbf {\bibinfo
  {volume} {54}},\ \bibinfo {pages} {11169} (\bibinfo {year}
  {1996})}\BibitemShut {NoStop}%
\bibitem [{\citenamefont {Kresse}\ and\ \citenamefont
  {Joubert}(1999)}]{Kresse1999}%
  \BibitemOpen
  \bibfield  {author} {\bibinfo {author} {\bibfnamefont {G.}~\bibnamefont
  {Kresse}}\ and\ \bibinfo {author} {\bibfnamefont {D.}~\bibnamefont
  {Joubert}},\ }\href {\doibase 10.1103/PhysRevB.59.1758} {\bibfield  {journal}
  {\bibinfo  {journal} {Phys. Rev. B}\ }\textbf {\bibinfo {volume} {59}},\
  \bibinfo {pages} {1758} (\bibinfo {year} {1999})}\BibitemShut {NoStop}%
\bibitem [{\citenamefont {Bl\"ochl}(1994)}]{Blochl1994}%
  \BibitemOpen
  \bibfield  {author} {\bibinfo {author} {\bibfnamefont {P.~E.}\ \bibnamefont
  {Bl\"ochl}},\ }\href {\doibase 10.1103/PhysRevB.50.17953} {\bibfield
  {journal} {\bibinfo  {journal} {Phys. Rev. B}\ }\textbf {\bibinfo {volume}
  {50}},\ \bibinfo {pages} {17953} (\bibinfo {year} {1994})}\BibitemShut
  {NoStop}%
\bibitem [{\citenamefont {Perdew}\ \emph {et~al.}(2008)\citenamefont {Perdew},
  \citenamefont {Ruzsinszky}, \citenamefont {Csonka}, \citenamefont {Vydrov},
  \citenamefont {Scuseria}, \citenamefont {Constantin}, \citenamefont {Zhou},\
  and\ \citenamefont {Burke}}]{Perdew2008}%
  \BibitemOpen
  \bibfield  {author} {\bibinfo {author} {\bibfnamefont {J.~P.}\ \bibnamefont
  {Perdew}}, \bibinfo {author} {\bibfnamefont {A.}~\bibnamefont {Ruzsinszky}},
  \bibinfo {author} {\bibfnamefont {G.~I.}\ \bibnamefont {Csonka}}, \bibinfo
  {author} {\bibfnamefont {O.~A.}\ \bibnamefont {Vydrov}}, \bibinfo {author}
  {\bibfnamefont {G.~E.}\ \bibnamefont {Scuseria}}, \bibinfo {author}
  {\bibfnamefont {L.~A.}\ \bibnamefont {Constantin}}, \bibinfo {author}
  {\bibfnamefont {X.}~\bibnamefont {Zhou}}, \ and\ \bibinfo {author}
  {\bibfnamefont {K.}~\bibnamefont {Burke}},\ }\href {\doibase
  10.1103/PhysRevLett.100.136406} {\bibfield  {journal} {\bibinfo  {journal}
  {Phys. Rev. Lett.}\ }\textbf {\bibinfo {volume} {100}},\ \bibinfo {pages}
  {136406} (\bibinfo {year} {2008})}\BibitemShut {NoStop}%
\bibitem [{\citenamefont {Charles}\ and\ \citenamefont
  {Rondinelli}(2016)}]{PhysRevB.94.174108}%
  \BibitemOpen
  \bibfield  {author} {\bibinfo {author} {\bibfnamefont {N.}~\bibnamefont
  {Charles}}\ and\ \bibinfo {author} {\bibfnamefont {J.~M.}\ \bibnamefont
  {Rondinelli}},\ }\href {\doibase 10.1103/PhysRevB.94.174108} {\bibfield
  {journal} {\bibinfo  {journal} {Phys. Rev. B}\ }\textbf {\bibinfo {volume}
  {94}},\ \bibinfo {pages} {174108} (\bibinfo {year} {2016})}\BibitemShut
  {NoStop}%
\bibitem [{\citenamefont {Liechtenstein}\ \emph {et~al.}(1995)\citenamefont
  {Liechtenstein}, \citenamefont {Anisimov},\ and\ \citenamefont
  {Zaanen}}]{Liechtenstein1995}%
  \BibitemOpen
  \bibfield  {author} {\bibinfo {author} {\bibfnamefont {A.~I.}\ \bibnamefont
  {Liechtenstein}}, \bibinfo {author} {\bibfnamefont {V.~I.}\ \bibnamefont
  {Anisimov}}, \ and\ \bibinfo {author} {\bibfnamefont {J.}~\bibnamefont
  {Zaanen}},\ }\href {\doibase 10.1103/PhysRevB.52.R5467} {\bibfield  {journal}
  {\bibinfo  {journal} {Phys. Rev. B}\ }\textbf {\bibinfo {volume} {52}},\
  \bibinfo {pages} {R5467} (\bibinfo {year} {1995})}\BibitemShut {NoStop}%
\bibitem [{\citenamefont {Heyd}\ \emph {et~al.}(2003)\citenamefont {Heyd},
  \citenamefont {Scuseria},\ and\ \citenamefont {Ernzerhof}}]{HSE}%
  \BibitemOpen
  \bibfield  {author} {\bibinfo {author} {\bibfnamefont {J.}~\bibnamefont
  {Heyd}}, \bibinfo {author} {\bibfnamefont {G.~E.}\ \bibnamefont {Scuseria}},
  \ and\ \bibinfo {author} {\bibfnamefont {M.}~\bibnamefont {Ernzerhof}},\
  }\href {\doibase 10.1063/1.1564060} {\bibfield  {journal} {\bibinfo
  {journal} {The Journal of Chemical Physics}\ }\textbf {\bibinfo {volume}
  {118}},\ \bibinfo {pages} {8207} (\bibinfo {year} {2003})}\BibitemShut
  {NoStop}%
\bibitem [{\citenamefont {Krukau}\ \emph {et~al.}(2006)\citenamefont {Krukau},
  \citenamefont {Vydrov}, \citenamefont {Izmaylov},\ and\ \citenamefont
  {Scuseria}}]{HSE06}%
  \BibitemOpen
  \bibfield  {author} {\bibinfo {author} {\bibfnamefont {A.~V.}\ \bibnamefont
  {Krukau}}, \bibinfo {author} {\bibfnamefont {O.~A.}\ \bibnamefont {Vydrov}},
  \bibinfo {author} {\bibfnamefont {A.~F.}\ \bibnamefont {Izmaylov}}, \ and\
  \bibinfo {author} {\bibfnamefont {G.~E.}\ \bibnamefont {Scuseria}},\ }\href
  {\doibase 10.1063/1.2404663} {\bibfield  {journal} {\bibinfo  {journal} {The
  Journal of Chemical Physics}\ }\textbf {\bibinfo {volume} {125}},\ \bibinfo
  {eid} {224106} (\bibinfo {year} {2006})}\BibitemShut {NoStop}%
\bibitem [{\citenamefont {Hobbs}\ \emph {et~al.}(2000)\citenamefont {Hobbs},
  \citenamefont {Kresse},\ and\ \citenamefont {Hafner}}]{Hobbs2000}%
  \BibitemOpen
  \bibfield  {author} {\bibinfo {author} {\bibfnamefont {D.}~\bibnamefont
  {Hobbs}}, \bibinfo {author} {\bibfnamefont {G.}~\bibnamefont {Kresse}}, \
  and\ \bibinfo {author} {\bibfnamefont {J.}~\bibnamefont {Hafner}},\ }\href
  {\doibase 10.1103/PhysRevB.62.11556} {\bibfield  {journal} {\bibinfo
  {journal} {Phys. Rev. B}\ }\textbf {\bibinfo {volume} {62}},\ \bibinfo
  {pages} {11556} (\bibinfo {year} {2000})}\BibitemShut {NoStop}%
\bibitem [{\citenamefont {Gonze}\ and\ \citenamefont {Lee}(1997)}]{gonze1997}%
  \BibitemOpen
  \bibfield  {author} {\bibinfo {author} {\bibfnamefont {X.}~\bibnamefont
  {Gonze}}\ and\ \bibinfo {author} {\bibfnamefont {C.}~\bibnamefont {Lee}},\
  }\href {\doibase 10.1103/PhysRevB.55.10355} {\bibfield  {journal} {\bibinfo
  {journal} {Phys. Rev. B}\ }\textbf {\bibinfo {volume} {55}},\ \bibinfo
  {pages} {10355} (\bibinfo {year} {1997})}\BibitemShut {NoStop}%
\bibitem [{\citenamefont {Togo}\ and\ \citenamefont {Tanaka}(2015)}]{phonopy}%
  \BibitemOpen
  \bibfield  {author} {\bibinfo {author} {\bibfnamefont {A.}~\bibnamefont
  {Togo}}\ and\ \bibinfo {author} {\bibfnamefont {I.}~\bibnamefont {Tanaka}},\
  }\href@noop {} {\bibfield  {journal} {\bibinfo  {journal} {Scr. Mater.}\
  }\textbf {\bibinfo {volume} {108}},\ \bibinfo {pages} {1} (\bibinfo {year}
  {2015})}\BibitemShut {NoStop}%
\bibitem [{\citenamefont {Roberto}\ \emph {et~al.}(2005)\citenamefont
  {Roberto}, \citenamefont {Roberto}, \citenamefont {Bartolomeo}, \citenamefont
  {Carla}, \citenamefont {R.},\ and\ \citenamefont {M.}}]{Roberto2005}%
  \BibitemOpen
  \bibfield  {author} {\bibinfo {author} {\bibfnamefont {D.}~\bibnamefont
  {Roberto}}, \bibinfo {author} {\bibfnamefont {O.}~\bibnamefont {Roberto}},
  \bibinfo {author} {\bibfnamefont {C.}~\bibnamefont {Bartolomeo}}, \bibinfo
  {author} {\bibfnamefont {R.}~\bibnamefont {Carla}}, \bibinfo {author}
  {\bibfnamefont {S.~V.}\ \bibnamefont {R.}}, \ and\ \bibinfo {author}
  {\bibfnamefont {Z.-W.~C.}\ \bibnamefont {M.}},\ }\href {\doibase
  10.1524/zkri.220.5.571.65065} {\bibfield  {journal} {\bibinfo  {journal}
  {Zeitschrift f�r Kristallographie - Crystalline Materials}\ }\textbf
  {\bibinfo {volume} {220}},\ \bibinfo {pages} {571} (\bibinfo {year}
  {2005})}\BibitemShut {NoStop}%
\bibitem [{\citenamefont {Bilc}\ \emph {et~al.}(2008)\citenamefont {Bilc},
  \citenamefont {Orlando}, \citenamefont {Shaltaf}, \citenamefont {Rignanese},
  \citenamefont {\'I\~niguez},\ and\ \citenamefont
  {Ghosez}}]{PhysRevB.77.165107}%
  \BibitemOpen
  \bibfield  {author} {\bibinfo {author} {\bibfnamefont {D.~I.}\ \bibnamefont
  {Bilc}}, \bibinfo {author} {\bibfnamefont {R.}~\bibnamefont {Orlando}},
  \bibinfo {author} {\bibfnamefont {R.}~\bibnamefont {Shaltaf}}, \bibinfo
  {author} {\bibfnamefont {G.-M.}\ \bibnamefont {Rignanese}}, \bibinfo {author}
  {\bibfnamefont {J.}~\bibnamefont {\'I\~niguez}}, \ and\ \bibinfo {author}
  {\bibfnamefont {P.}~\bibnamefont {Ghosez}},\ }\href {\doibase
  10.1103/PhysRevB.77.165107} {\bibfield  {journal} {\bibinfo  {journal} {Phys.
  Rev. B}\ }\textbf {\bibinfo {volume} {77}},\ \bibinfo {pages} {165107}
  (\bibinfo {year} {2008})}\BibitemShut {NoStop}%
\bibitem [{\citenamefont
  {Becke}(1993)}]{:/content/aip/journal/jcp/98/2/10.1063/1.464304}%
  \BibitemOpen
  \bibfield  {author} {\bibinfo {author} {\bibfnamefont {A.~D.}\ \bibnamefont
  {Becke}},\ }\href {\doibase 10.1063/1.464304} {\bibfield  {journal} {\bibinfo
   {journal} {The Journal of Chemical Physics}\ }\textbf {\bibinfo {volume}
  {98}},\ \bibinfo {pages} {1372} (\bibinfo {year} {1993})}\BibitemShut
  {NoStop}%
\bibitem [{\citenamefont
  {Becke}(1992)}]{:/content/aip/journal/jcp/96/3/10.1063/1.462066}%
  \BibitemOpen
  \bibfield  {author} {\bibinfo {author} {\bibfnamefont {A.~D.}\ \bibnamefont
  {Becke}},\ }\href {\doibase 10.1063/1.462066} {\bibfield  {journal} {\bibinfo
   {journal} {The Journal of Chemical Physics}\ }\textbf {\bibinfo {volume}
  {96}},\ \bibinfo {pages} {2155} (\bibinfo {year} {1992})}\BibitemShut
  {NoStop}%
\bibitem [{\citenamefont {Perdew}\ \emph {et~al.}(1996)\citenamefont {Perdew},
  \citenamefont {Ernzerhof},\ and\ \citenamefont {Burke}}]{PBE0}%
  \BibitemOpen
  \bibfield  {author} {\bibinfo {author} {\bibfnamefont {J.~P.}\ \bibnamefont
  {Perdew}}, \bibinfo {author} {\bibfnamefont {M.}~\bibnamefont {Ernzerhof}}, \
  and\ \bibinfo {author} {\bibfnamefont {K.}~\bibnamefont {Burke}},\ }\href
  {\doibase 10.1063/1.472933} {\bibfield  {journal} {\bibinfo  {journal} {The
  Journal of Chemical Physics}\ }\textbf {\bibinfo {volume} {105}},\ \bibinfo
  {pages} {9982} (\bibinfo {year} {1996})}\BibitemShut {NoStop}%
\bibitem [{\citenamefont {Madelung}\ \emph {et~al.}(2000)\citenamefont
  {Madelung}, \citenamefont {R{\"o}ssler},\ and\ \citenamefont
  {Schulz}}]{ref1}%
  \BibitemOpen
  \bibinfo {editor} {\bibfnamefont {O.}~\bibnamefont {Madelung}}, \bibinfo
  {editor} {\bibfnamefont {U.}~\bibnamefont {R{\"o}ssler}}, \ and\ \bibinfo
  {editor} {\bibfnamefont {M.}~\bibnamefont {Schulz}},\ eds.,\ \enquote
  {\bibinfo {title} {${\text{ktao}}_{3}$ crystal structure, lattice parameters,
  thermal expansion},}\ in\ \href {\doibase 10.1007/10717201_192} {\emph
  {\bibinfo {booktitle} {Ternary Compounds, Organic Semiconductors}}}\
  (\bibinfo  {publisher} {Springer Berlin Heidelberg},\ \bibinfo {address}
  {Berlin, Heidelberg},\ \bibinfo {year} {2000})\ pp.\ \bibinfo {pages}
  {1--3}\BibitemShut {NoStop}%
\bibitem [{\citenamefont {Knox}(1961)}]{Knox1961}%
  \BibitemOpen
  \bibfield  {author} {\bibinfo {author} {\bibfnamefont {K.}~\bibnamefont
  {Knox}},\ }\href {\doibase 10.1107/S0365110X61001868} {\bibfield  {journal}
  {\bibinfo  {journal} {Acta Crystallographica}\ }\textbf {\bibinfo {volume}
  {14}},\ \bibinfo {pages} {583} (\bibinfo {year} {1961})}\BibitemShut
  {NoStop}%
\bibitem [{\citenamefont {Okazaki}\ \emph {et~al.}(1959)\citenamefont
  {Okazaki}, \citenamefont {Suemune},\ and\ \citenamefont
  {Fuchikami}}]{doi:10.1143/JPSJ.14.1823}%
  \BibitemOpen
  \bibfield  {author} {\bibinfo {author} {\bibfnamefont {A.}~\bibnamefont
  {Okazaki}}, \bibinfo {author} {\bibfnamefont {Y.}~\bibnamefont {Suemune}}, \
  and\ \bibinfo {author} {\bibfnamefont {T.}~\bibnamefont {Fuchikami}},\ }\href
  {\doibase 10.1143/JPSJ.14.1823} {\bibfield  {journal} {\bibinfo  {journal}
  {Journal of the Physical Society of Japan}\ }\textbf {\bibinfo {volume}
  {14}},\ \bibinfo {pages} {1823} (\bibinfo {year} {1959})}\BibitemShut
  {NoStop}%
\bibitem [{\citenamefont {Pari}\ \emph {et~al.}(1994)\citenamefont {Pari},
  \citenamefont {Mathi~Jaya},\ and\ \citenamefont
  {Asokamani}}]{PhysRevB.50.8166}%
  \BibitemOpen
  \bibfield  {author} {\bibinfo {author} {\bibfnamefont {G.}~\bibnamefont
  {Pari}}, \bibinfo {author} {\bibfnamefont {S.}~\bibnamefont {Mathi~Jaya}}, \
  and\ \bibinfo {author} {\bibfnamefont {R.}~\bibnamefont {Asokamani}},\ }\href
  {\doibase 10.1103/PhysRevB.50.8166} {\bibfield  {journal} {\bibinfo
  {journal} {Phys. Rev. B}\ }\textbf {\bibinfo {volume} {50}},\ \bibinfo
  {pages} {8166} (\bibinfo {year} {1994})}\BibitemShut {NoStop}%
\bibitem [{\citenamefont {Bristowe}\ \emph {et~al.}(2009)\citenamefont
  {Bristowe}, \citenamefont {Artacho},\ and\ \citenamefont
  {Littlewood}}]{PhysRevB.80.045425}%
  \BibitemOpen
  \bibfield  {author} {\bibinfo {author} {\bibfnamefont {N.~C.}\ \bibnamefont
  {Bristowe}}, \bibinfo {author} {\bibfnamefont {E.}~\bibnamefont {Artacho}}, \
  and\ \bibinfo {author} {\bibfnamefont {P.~B.}\ \bibnamefont {Littlewood}},\
  }\href {\doibase 10.1103/PhysRevB.80.045425} {\bibfield  {journal} {\bibinfo
  {journal} {Phys. Rev. B}\ }\textbf {\bibinfo {volume} {80}},\ \bibinfo
  {pages} {045425} (\bibinfo {year} {2009})}\BibitemShut {NoStop}%
\bibitem [{\citenamefont {Jellison}\ \emph {et~al.}(2006)\citenamefont
  {Jellison}, \citenamefont {Paulauskas}, \citenamefont {Boatner},\ and\
  \citenamefont {Singh}}]{PhysRevB.74.155130}%
  \BibitemOpen
  \bibfield  {author} {\bibinfo {author} {\bibfnamefont {G.~E.}\ \bibnamefont
  {Jellison}}, \bibinfo {author} {\bibfnamefont {I.}~\bibnamefont
  {Paulauskas}}, \bibinfo {author} {\bibfnamefont {L.~A.}\ \bibnamefont
  {Boatner}}, \ and\ \bibinfo {author} {\bibfnamefont {D.~J.}\ \bibnamefont
  {Singh}},\ }\href {\doibase 10.1103/PhysRevB.74.155130} {\bibfield  {journal}
  {\bibinfo  {journal} {Phys. Rev. B}\ }\textbf {\bibinfo {volume} {74}},\
  \bibinfo {pages} {155130} (\bibinfo {year} {2006})}\BibitemShut {NoStop}%
\bibitem [{\citenamefont {Tyunina}\ \emph {et~al.}(2010)\citenamefont
  {Tyunina}, \citenamefont {Narkilahti}, \citenamefont {Plekh}, \citenamefont
  {Oja}, \citenamefont {Nieminen}, \citenamefont {Dejneka},\ and\ \citenamefont
  {Trepakov}}]{PhysRevLett.104.227601}%
  \BibitemOpen
  \bibfield  {author} {\bibinfo {author} {\bibfnamefont {M.}~\bibnamefont
  {Tyunina}}, \bibinfo {author} {\bibfnamefont {J.}~\bibnamefont {Narkilahti}},
  \bibinfo {author} {\bibfnamefont {M.}~\bibnamefont {Plekh}}, \bibinfo
  {author} {\bibfnamefont {R.}~\bibnamefont {Oja}}, \bibinfo {author}
  {\bibfnamefont {R.~M.}\ \bibnamefont {Nieminen}}, \bibinfo {author}
  {\bibfnamefont {A.}~\bibnamefont {Dejneka}}, \ and\ \bibinfo {author}
  {\bibfnamefont {V.}~\bibnamefont {Trepakov}},\ }\href {\doibase
  10.1103/PhysRevLett.104.227601} {\bibfield  {journal} {\bibinfo  {journal}
  {Phys. Rev. Lett.}\ }\textbf {\bibinfo {volume} {104}},\ \bibinfo {pages}
  {227601} (\bibinfo {year} {2010})}\BibitemShut {NoStop}%
\bibitem [{\citenamefont {Vanderbilt}(2000)}]{Vanderbilt2000147}%
  \BibitemOpen
  \bibfield  {author} {\bibinfo {author} {\bibfnamefont {D.}~\bibnamefont
  {Vanderbilt}},\ }\href {\doibase
  http://dx.doi.org/10.1016/S0022-3697(99)00273-5} {\bibfield  {journal}
  {\bibinfo  {journal} {Journal of Physics and Chemistry of Solids}\ }\textbf
  {\bibinfo {volume} {61}},\ \bibinfo {pages} {147 } (\bibinfo {year}
  {2000})}\BibitemShut {NoStop}%
\bibitem [{\citenamefont {Cancellieri}\ \emph {et~al.}(2011)\citenamefont
  {Cancellieri}, \citenamefont {Fontaine}, \citenamefont {Gariglio},
  \citenamefont {Reyren}, \citenamefont {Caviglia}, \citenamefont {F\^ete},
  \citenamefont {Leake}, \citenamefont {Pauli}, \citenamefont {Willmott},
  \citenamefont {Stengel}, \citenamefont {Ghosez},\ and\ \citenamefont
  {Triscone}}]{PhysRevLett.107.056102}%
  \BibitemOpen
  \bibfield  {author} {\bibinfo {author} {\bibfnamefont {C.}~\bibnamefont
  {Cancellieri}}, \bibinfo {author} {\bibfnamefont {D.}~\bibnamefont
  {Fontaine}}, \bibinfo {author} {\bibfnamefont {S.}~\bibnamefont {Gariglio}},
  \bibinfo {author} {\bibfnamefont {N.}~\bibnamefont {Reyren}}, \bibinfo
  {author} {\bibfnamefont {A.~D.}\ \bibnamefont {Caviglia}}, \bibinfo {author}
  {\bibfnamefont {A.}~\bibnamefont {F\^ete}}, \bibinfo {author} {\bibfnamefont
  {S.~J.}\ \bibnamefont {Leake}}, \bibinfo {author} {\bibfnamefont {S.~A.}\
  \bibnamefont {Pauli}}, \bibinfo {author} {\bibfnamefont {P.~R.}\ \bibnamefont
  {Willmott}}, \bibinfo {author} {\bibfnamefont {M.}~\bibnamefont {Stengel}},
  \bibinfo {author} {\bibfnamefont {P.}~\bibnamefont {Ghosez}}, \ and\ \bibinfo
  {author} {\bibfnamefont {J.-M.}\ \bibnamefont {Triscone}},\ }\href {\doibase
  10.1103/PhysRevLett.107.056102} {\bibfield  {journal} {\bibinfo  {journal}
  {Phys. Rev. Lett.}\ }\textbf {\bibinfo {volume} {107}},\ \bibinfo {pages}
  {056102} (\bibinfo {year} {2011})}\BibitemShut {NoStop}%
\bibitem [{\citenamefont {Bousquet}\ \emph {et~al.}(2010)\citenamefont
  {Bousquet}, \citenamefont {Junquera},\ and\ \citenamefont
  {Ghosez}}]{PhysRevB.82.045426}%
  \BibitemOpen
  \bibfield  {author} {\bibinfo {author} {\bibfnamefont {E.}~\bibnamefont
  {Bousquet}}, \bibinfo {author} {\bibfnamefont {J.}~\bibnamefont {Junquera}},
  \ and\ \bibinfo {author} {\bibfnamefont {P.}~\bibnamefont {Ghosez}},\ }\href
  {\doibase 10.1103/PhysRevB.82.045426} {\bibfield  {journal} {\bibinfo
  {journal} {Phys. Rev. B}\ }\textbf {\bibinfo {volume} {82}},\ \bibinfo
  {pages} {045426} (\bibinfo {year} {2010})}\BibitemShut {NoStop}%
\bibitem [{\citenamefont {Yu}\ \emph {et~al.}(2012)\citenamefont {Yu},
  \citenamefont {Chu},\ and\ \citenamefont {Ramesh}}]{Yu2012}%
  \BibitemOpen
  \bibfield  {author} {\bibinfo {author} {\bibfnamefont {P.}~\bibnamefont
  {Yu}}, \bibinfo {author} {\bibfnamefont {Y.}~\bibnamefont {Chu}}, \ and\
  \bibinfo {author} {\bibfnamefont {R.}~\bibnamefont {Ramesh}},\ }\href
  {http://www.sciencedirect.com/science/article/pii/S1369702112701372}
  {\bibfield  {journal} {\bibinfo  {journal} {Materials Today}\ }\textbf
  {\bibinfo {volume} {15}},\ \bibinfo {pages} {320} (\bibinfo {year}
  {2012})}\BibitemShut {NoStop}%
\bibitem [{\citenamefont {Bychkov}\ and\ \citenamefont
  {Rashba}(1984)}]{rashba1984}%
  \BibitemOpen
  \bibfield  {author} {\bibinfo {author} {\bibfnamefont {Y.~A.}\ \bibnamefont
  {Bychkov}}\ and\ \bibinfo {author} {\bibfnamefont {E.~I.}\ \bibnamefont
  {Rashba}},\ }\href {http://stacks.iop.org/0022-3719/17/i=33/a=015} {\bibfield
   {journal} {\bibinfo  {journal} {Journal of Physics C: Solid State Physics}\
  }\textbf {\bibinfo {volume} {17}},\ \bibinfo {pages} {6039} (\bibinfo {year}
  {1984})}\BibitemShut {NoStop}%
\bibitem [{\citenamefont {Caviglia}\ \emph {et~al.}(2010)\citenamefont
  {Caviglia}, \citenamefont {Gabay}, \citenamefont {Gariglio}, \citenamefont
  {Reyren}, \citenamefont {Cancellieri},\ and\ \citenamefont
  {Triscone}}]{PhysRevLett.104.126803}%
  \BibitemOpen
  \bibfield  {author} {\bibinfo {author} {\bibfnamefont {A.~D.}\ \bibnamefont
  {Caviglia}}, \bibinfo {author} {\bibfnamefont {M.}~\bibnamefont {Gabay}},
  \bibinfo {author} {\bibfnamefont {S.}~\bibnamefont {Gariglio}}, \bibinfo
  {author} {\bibfnamefont {N.}~\bibnamefont {Reyren}}, \bibinfo {author}
  {\bibfnamefont {C.}~\bibnamefont {Cancellieri}}, \ and\ \bibinfo {author}
  {\bibfnamefont {J.-M.}\ \bibnamefont {Triscone}},\ }\href {\doibase
  10.1103/PhysRevLett.104.126803} {\bibfield  {journal} {\bibinfo  {journal}
  {Phys. Rev. Lett.}\ }\textbf {\bibinfo {volume} {104}},\ \bibinfo {pages}
  {126803} (\bibinfo {year} {2010})}\BibitemShut {NoStop}%
\bibitem [{\citenamefont {Zhong}\ \emph {et~al.}(2013)\citenamefont {Zhong},
  \citenamefont {T\'oth},\ and\ \citenamefont {Held}}]{PhysRevB.87.161102}%
  \BibitemOpen
  \bibfield  {author} {\bibinfo {author} {\bibfnamefont {Z.}~\bibnamefont
  {Zhong}}, \bibinfo {author} {\bibfnamefont {A.}~\bibnamefont {T\'oth}}, \
  and\ \bibinfo {author} {\bibfnamefont {K.}~\bibnamefont {Held}},\ }\href
  {\doibase 10.1103/PhysRevB.87.161102} {\bibfield  {journal} {\bibinfo
  {journal} {Phys. Rev. B}\ }\textbf {\bibinfo {volume} {87}},\ \bibinfo
  {pages} {161102} (\bibinfo {year} {2013})}\BibitemShut {NoStop}%
\bibitem [{\citenamefont {Hurand}\ \emph {et~al.}(2015)\citenamefont {Hurand},
  \citenamefont {Jouan}, \citenamefont {Feuillet-Palma}, \citenamefont {Singh},
  \citenamefont {Biscaras}, \citenamefont {Lesne}, \citenamefont {Reyren},
  \citenamefont {Barth{\'e}l{\'e}my}, \citenamefont {Bibes}, \citenamefont
  {Villegas}, \citenamefont {Ulysse}, \citenamefont {Lafosse}, \citenamefont
  {Pannetier-Lecoeur}, \citenamefont {Caprara}, \citenamefont {Grilli},
  \citenamefont {Lesueur},\ and\ \citenamefont {Bergeal}}]{Hurand2015}%
  \BibitemOpen
  \bibfield  {author} {\bibinfo {author} {\bibfnamefont {S.}~\bibnamefont
  {Hurand}}, \bibinfo {author} {\bibfnamefont {A.}~\bibnamefont {Jouan}},
  \bibinfo {author} {\bibfnamefont {C.}~\bibnamefont {Feuillet-Palma}},
  \bibinfo {author} {\bibfnamefont {G.}~\bibnamefont {Singh}}, \bibinfo
  {author} {\bibfnamefont {J.}~\bibnamefont {Biscaras}}, \bibinfo {author}
  {\bibfnamefont {E.}~\bibnamefont {Lesne}}, \bibinfo {author} {\bibfnamefont
  {N.}~\bibnamefont {Reyren}}, \bibinfo {author} {\bibfnamefont
  {A.}~\bibnamefont {Barth{\'e}l{\'e}my}}, \bibinfo {author} {\bibfnamefont
  {M.}~\bibnamefont {Bibes}}, \bibinfo {author} {\bibfnamefont {J.~E.}\
  \bibnamefont {Villegas}}, \bibinfo {author} {\bibfnamefont {C.}~\bibnamefont
  {Ulysse}}, \bibinfo {author} {\bibfnamefont {X.}~\bibnamefont {Lafosse}},
  \bibinfo {author} {\bibfnamefont {M.}~\bibnamefont {Pannetier-Lecoeur}},
  \bibinfo {author} {\bibfnamefont {S.}~\bibnamefont {Caprara}}, \bibinfo
  {author} {\bibfnamefont {M.}~\bibnamefont {Grilli}}, \bibinfo {author}
  {\bibfnamefont {J.}~\bibnamefont {Lesueur}}, \ and\ \bibinfo {author}
  {\bibfnamefont {N.}~\bibnamefont {Bergeal}},\ }\href
  {http://dx.doi.org/10.1038/srep12751} {\bibfield  {journal} {\bibinfo
  {journal} {Scientific Reports}\ }\textbf {\bibinfo {volume} {5}},\ \bibinfo
  {pages} {12751 EP } (\bibinfo {year} {2015})}\BibitemShut {NoStop}%
\bibitem [{\citenamefont {Nakamura}\ \emph {et~al.}(2012)\citenamefont
  {Nakamura}, \citenamefont {Koga},\ and\ \citenamefont
  {Kimura}}]{PhysRevLett.108.206601}%
  \BibitemOpen
  \bibfield  {author} {\bibinfo {author} {\bibfnamefont {H.}~\bibnamefont
  {Nakamura}}, \bibinfo {author} {\bibfnamefont {T.}~\bibnamefont {Koga}}, \
  and\ \bibinfo {author} {\bibfnamefont {T.}~\bibnamefont {Kimura}},\ }\href
  {\doibase 10.1103/PhysRevLett.108.206601} {\bibfield  {journal} {\bibinfo
  {journal} {Phys. Rev. Lett.}\ }\textbf {\bibinfo {volume} {108}},\ \bibinfo
  {pages} {206601} (\bibinfo {year} {2012})}\BibitemShut {NoStop}%
\bibitem [{\citenamefont {Nakamura}\ and\ \citenamefont
  {Kimura}(2009)}]{PhysRevB.80.121308}%
  \BibitemOpen
  \bibfield  {author} {\bibinfo {author} {\bibfnamefont {H.}~\bibnamefont
  {Nakamura}}\ and\ \bibinfo {author} {\bibfnamefont {T.}~\bibnamefont
  {Kimura}},\ }\href {\doibase 10.1103/PhysRevB.80.121308} {\bibfield
  {journal} {\bibinfo  {journal} {Phys. Rev. B}\ }\textbf {\bibinfo {volume}
  {80}},\ \bibinfo {pages} {121308} (\bibinfo {year} {2009})}\BibitemShut
  {NoStop}%
\bibitem [{\citenamefont {Lee}\ and\ \citenamefont
  {Choi}(2012)}]{PhysRevB.86.045437}%
  \BibitemOpen
  \bibfield  {author} {\bibinfo {author} {\bibfnamefont {H.}~\bibnamefont
  {Lee}}\ and\ \bibinfo {author} {\bibfnamefont {H.~J.}\ \bibnamefont {Choi}},\
  }\href {\doibase 10.1103/PhysRevB.86.045437} {\bibfield  {journal} {\bibinfo
  {journal} {Phys. Rev. B}\ }\textbf {\bibinfo {volume} {86}},\ \bibinfo
  {pages} {045437} (\bibinfo {year} {2012})}\BibitemShut {NoStop}%
\bibitem [{\citenamefont {Park}\ and\ \citenamefont {Kim}(2015)}]{Park20156}%
  \BibitemOpen
  \bibfield  {author} {\bibinfo {author} {\bibfnamefont {S.~R.}\ \bibnamefont
  {Park}}\ and\ \bibinfo {author} {\bibfnamefont {C.}~\bibnamefont {Kim}},\
  }\href {\doibase http://dx.doi.org/10.1016/j.elspec.2014.12.009} {\bibfield
  {journal} {\bibinfo  {journal} {Journal of Electron Spectroscopy and Related
  Phenomena}\ }\textbf {\bibinfo {volume} {201}},\ \bibinfo {pages} {6 }
  (\bibinfo {year} {2015})},\ \bibinfo {note} {special issue on electron
  spectroscopy for Rashba spin-orbit interaction}\BibitemShut {NoStop}%
\bibitem [{\citenamefont {Herath}\ \emph {et~al.}(2019)\citenamefont {Herath},
  \citenamefont {Tavadze}, \citenamefont {He}, \citenamefont {Bousquet},
  \citenamefont {Singh}, \citenamefont {Mu{\~n}oz},\ and\ \citenamefont
  {Romero}}]{herath2019pyprocar}%
  \BibitemOpen
  \bibfield  {author} {\bibinfo {author} {\bibfnamefont {U.}~\bibnamefont
  {Herath}}, \bibinfo {author} {\bibfnamefont {P.}~\bibnamefont {Tavadze}},
  \bibinfo {author} {\bibfnamefont {X.}~\bibnamefont {He}}, \bibinfo {author}
  {\bibfnamefont {E.}~\bibnamefont {Bousquet}}, \bibinfo {author}
  {\bibfnamefont {S.}~\bibnamefont {Singh}}, \bibinfo {author} {\bibfnamefont
  {F.}~\bibnamefont {Mu{\~n}oz}}, \ and\ \bibinfo {author} {\bibfnamefont
  {A.~H.}\ \bibnamefont {Romero}},\ }\href@noop {} {\bibfield  {journal}
  {\bibinfo  {journal} {arXiv preprint arXiv:1906.11387}\ } (\bibinfo {year}
  {2019})}\BibitemShut {NoStop}%
\bibitem [{\citenamefont {Shanavas}(2015)}]{Shanavas2014}%
  \BibitemOpen
  \bibfield  {author} {\bibinfo {author} {\bibfnamefont {K.}~\bibnamefont
  {Shanavas}},\ }\href {\doibase
  http://dx.doi.org/10.1016/j.elspec.2014.08.005} {\bibfield  {journal}
  {\bibinfo  {journal} {Journal of Electron Spectroscopy and Related
  Phenomena}\ }\textbf {\bibinfo {volume} {201}},\ \bibinfo {pages} {121 }
  (\bibinfo {year} {2015})}\BibitemShut {NoStop}%
\bibitem [{Note1()}]{Note1}%
  \BibitemOpen
  \bibinfo {note} {The linear-Rashba effect has the $E$ vs. $k$ form: $E^{\pm
  }$($k$) = (${\mathchar '26\mkern -9muh}^2$$k^2$/$2m^*$) $\pm $ $\alpha $|$k$|
  coming from the Hamiltonian of the form $H_{R}$ = $\alpha $$\protect \mathcal
  {E}_z$$i(k_{-} \sigma _{+} - k_{+}\sigma _{-})$}\BibitemShut {NoStop}%
\bibitem [{\citenamefont {Ishizaka}\ \emph {et~al.}(2011)\citenamefont
  {Ishizaka}, \citenamefont {Bahramy}, \citenamefont {Murakawa}, \citenamefont
  {Sakano}, \citenamefont {Shimojima}, \citenamefont {Sonobe}, \citenamefont
  {Koizumi}, \citenamefont {Shin}, \citenamefont {Miyahara}, \citenamefont
  {Kimura}, \citenamefont {Miyamoto}, \citenamefont {Okuda}, \citenamefont
  {Namatame}, \citenamefont {Taniguchi}, \citenamefont {Arita}, \citenamefont
  {Nagaosa}, \citenamefont {Kobayashi}, \citenamefont {Murakami}, \citenamefont
  {Kumai}, \citenamefont {Kaneko}, \citenamefont {Onose},\ and\ \citenamefont
  {Tokura}}]{Ishizaka2011}%
  \BibitemOpen
  \bibfield  {author} {\bibinfo {author} {\bibfnamefont {K.}~\bibnamefont
  {Ishizaka}}, \bibinfo {author} {\bibfnamefont {M.~S.}\ \bibnamefont
  {Bahramy}}, \bibinfo {author} {\bibfnamefont {H.}~\bibnamefont {Murakawa}},
  \bibinfo {author} {\bibfnamefont {M.}~\bibnamefont {Sakano}}, \bibinfo
  {author} {\bibfnamefont {T.}~\bibnamefont {Shimojima}}, \bibinfo {author}
  {\bibfnamefont {T.}~\bibnamefont {Sonobe}}, \bibinfo {author} {\bibfnamefont
  {K.}~\bibnamefont {Koizumi}}, \bibinfo {author} {\bibfnamefont
  {S.}~\bibnamefont {Shin}}, \bibinfo {author} {\bibfnamefont {H.}~\bibnamefont
  {Miyahara}}, \bibinfo {author} {\bibfnamefont {A.}~\bibnamefont {Kimura}},
  \bibinfo {author} {\bibfnamefont {K.}~\bibnamefont {Miyamoto}}, \bibinfo
  {author} {\bibfnamefont {T.}~\bibnamefont {Okuda}}, \bibinfo {author}
  {\bibfnamefont {H.}~\bibnamefont {Namatame}}, \bibinfo {author}
  {\bibfnamefont {M.}~\bibnamefont {Taniguchi}}, \bibinfo {author}
  {\bibfnamefont {R.}~\bibnamefont {Arita}}, \bibinfo {author} {\bibfnamefont
  {N.}~\bibnamefont {Nagaosa}}, \bibinfo {author} {\bibfnamefont
  {K.}~\bibnamefont {Kobayashi}}, \bibinfo {author} {\bibfnamefont
  {Y.}~\bibnamefont {Murakami}}, \bibinfo {author} {\bibfnamefont
  {R.}~\bibnamefont {Kumai}}, \bibinfo {author} {\bibfnamefont
  {Y.}~\bibnamefont {Kaneko}}, \bibinfo {author} {\bibfnamefont
  {Y.}~\bibnamefont {Onose}}, \ and\ \bibinfo {author} {\bibfnamefont
  {Y.}~\bibnamefont {Tokura}},\ }\href {\doibase 10.1038/nmat3051} {\bibfield
  {journal} {\bibinfo  {journal} {Nat Mater}\ }\textbf {\bibinfo {volume}
  {10}},\ \bibinfo {pages} {521} (\bibinfo {year} {2011})}\BibitemShut
  {NoStop}%
\bibitem [{\citenamefont {King}\ \emph {et~al.}(2012)\citenamefont {King},
  \citenamefont {He}, \citenamefont {Eknapakul}, \citenamefont {Buaphet},
  \citenamefont {Mo}, \citenamefont {Kaneko}, \citenamefont {Harashima},
  \citenamefont {Hikita}, \citenamefont {Bahramy}, \citenamefont {Bell},
  \citenamefont {Hussain}, \citenamefont {Tokura}, \citenamefont {Shen},
  \citenamefont {Hwang}, \citenamefont {Baumberger},\ and\ \citenamefont
  {Meevasana}}]{PhysRevLett.108.117602}%
  \BibitemOpen
  \bibfield  {author} {\bibinfo {author} {\bibfnamefont {P.~D.~C.}\
  \bibnamefont {King}}, \bibinfo {author} {\bibfnamefont {R.~H.}\ \bibnamefont
  {He}}, \bibinfo {author} {\bibfnamefont {T.}~\bibnamefont {Eknapakul}},
  \bibinfo {author} {\bibfnamefont {P.}~\bibnamefont {Buaphet}}, \bibinfo
  {author} {\bibfnamefont {S.-K.}\ \bibnamefont {Mo}}, \bibinfo {author}
  {\bibfnamefont {Y.}~\bibnamefont {Kaneko}}, \bibinfo {author} {\bibfnamefont
  {S.}~\bibnamefont {Harashima}}, \bibinfo {author} {\bibfnamefont
  {Y.}~\bibnamefont {Hikita}}, \bibinfo {author} {\bibfnamefont {M.~S.}\
  \bibnamefont {Bahramy}}, \bibinfo {author} {\bibfnamefont {C.}~\bibnamefont
  {Bell}}, \bibinfo {author} {\bibfnamefont {Z.}~\bibnamefont {Hussain}},
  \bibinfo {author} {\bibfnamefont {Y.}~\bibnamefont {Tokura}}, \bibinfo
  {author} {\bibfnamefont {Z.-X.}\ \bibnamefont {Shen}}, \bibinfo {author}
  {\bibfnamefont {H.~Y.}\ \bibnamefont {Hwang}}, \bibinfo {author}
  {\bibfnamefont {F.}~\bibnamefont {Baumberger}}, \ and\ \bibinfo {author}
  {\bibfnamefont {W.}~\bibnamefont {Meevasana}},\ }\href {\doibase
  10.1103/PhysRevLett.108.117602} {\bibfield  {journal} {\bibinfo  {journal}
  {Phys. Rev. Lett.}\ }\textbf {\bibinfo {volume} {108}},\ \bibinfo {pages}
  {117602} (\bibinfo {year} {2012})}\BibitemShut {NoStop}%
\bibitem [{\citenamefont {Bleibaum}\ and\ \citenamefont
  {Wachsmuth}(2006)}]{PhysRevB.74.195330}%
  \BibitemOpen
  \bibfield  {author} {\bibinfo {author} {\bibfnamefont {O.}~\bibnamefont
  {Bleibaum}}\ and\ \bibinfo {author} {\bibfnamefont {S.}~\bibnamefont
  {Wachsmuth}},\ }\href {\doibase 10.1103/PhysRevB.74.195330} {\bibfield
  {journal} {\bibinfo  {journal} {Phys. Rev. B}\ }\textbf {\bibinfo {volume}
  {74}},\ \bibinfo {pages} {195330} (\bibinfo {year} {2006})}\BibitemShut
  {NoStop}%
\bibitem [{\citenamefont {Moriya}\ \emph {et~al.}(2014)\citenamefont {Moriya},
  \citenamefont {Sawano}, \citenamefont {Hoshi}, \citenamefont {Masubuchi},
  \citenamefont {Shiraki}, \citenamefont {Wild}, \citenamefont {Neumann},
  \citenamefont {Abstreiter}, \citenamefont {Bougeard}, \citenamefont {Koga},\
  and\ \citenamefont {Machida}}]{PhysRevLett.113.086601}%
  \BibitemOpen
  \bibfield  {author} {\bibinfo {author} {\bibfnamefont {R.}~\bibnamefont
  {Moriya}}, \bibinfo {author} {\bibfnamefont {K.}~\bibnamefont {Sawano}},
  \bibinfo {author} {\bibfnamefont {Y.}~\bibnamefont {Hoshi}}, \bibinfo
  {author} {\bibfnamefont {S.}~\bibnamefont {Masubuchi}}, \bibinfo {author}
  {\bibfnamefont {Y.}~\bibnamefont {Shiraki}}, \bibinfo {author} {\bibfnamefont
  {A.}~\bibnamefont {Wild}}, \bibinfo {author} {\bibfnamefont {C.}~\bibnamefont
  {Neumann}}, \bibinfo {author} {\bibfnamefont {G.}~\bibnamefont {Abstreiter}},
  \bibinfo {author} {\bibfnamefont {D.}~\bibnamefont {Bougeard}}, \bibinfo
  {author} {\bibfnamefont {T.}~\bibnamefont {Koga}}, \ and\ \bibinfo {author}
  {\bibfnamefont {T.}~\bibnamefont {Machida}},\ }\href {\doibase
  10.1103/PhysRevLett.113.086601} {\bibfield  {journal} {\bibinfo  {journal}
  {Phys. Rev. Lett.}\ }\textbf {\bibinfo {volume} {113}},\ \bibinfo {pages}
  {086601} (\bibinfo {year} {2014})}\BibitemShut {NoStop}%
\end{thebibliography}%

\end{document}